\RequirePackage{fix-cm}
\documentclass[twocolumn]{svjour3}          % twocolumn
\smartqed  % flush right qed marks, e.g. at end of proof
\usepackage{graphicx}

\newcommand{\be}{\begin{equation}}
\newcommand{\qe}{\end{equation}}

\newcommand{\ba}{\begin{eqnarray}}
\newcommand{\ea}{\end{eqnarray}}

\usepackage{graphicx}
\usepackage[title]{appendix}

\usepackage{tikz-cd}
\usepackage{tikz}

\usetikzlibrary{decorations.markings}
\usetikzlibrary{shapes.geometric, arrows}
\usepackage{tikzscale}
\usepackage{filecontents}
\usepackage{wrapfig}
\usepackage{tcolorbox}
\usepackage{color}
\definecolor{blue1}{rgb}{0.0, 0.0, 1.0}
\definecolor{gray}{rgb}{0.9,0.9,0.9}
\definecolor{gray1}{rgb}{0.7,0.7,0.7}
\definecolor{gray2}{rgb}{0.8,0.8,0.8}
\definecolor{magenta}{rgb}{1.0, 0.0, 1.0}

\usepackage{hyperref}
\hypersetup{ colorlinks=true, urlcolor  = blue, linkcolor = blue,
citecolor = blue1,}

\usepackage{float}
\usepackage{amsmath,mathtools,amssymb,amsfonts}
\usepackage[scaled=.95]{helvet}

 \journalname{Noname}
\begin{document}
\title{Chaotic Synchronization of memristive neurons: Lyapunov function versus Hamilton function}


\author{Marius E. Yamakou 
}


\institute{M. E. Yamakou \at Angewandte Analysis, Department Mathematik\\ Friedrich-Alexander-Universit\"{a}t Erlangen-N\"{u}rnberg\\ Cauerstr. 11,
91058 Erlangen, Germany\\
             \email{marius.yamakou@fau.de} %\emph{Corresponding author}\\
                      %  \\
                      %  \\
            %of F. Author  %  if needed
}

\date{Received: date }%/ Accepted: date}

\maketitle

\begin{abstract}
We consider a physiologically improved version of the Hindmarsh-Rose neuron model.
In contrast to the standard 3D Hindmarsh-Rose model, the improved 5D version considers not only the calcium exchange between intracellular 
warehouse and the cytoplasm of the nerve cell, but also the magnetic flux of the electromagnetic field induced by the 
movement of ions across the membrane, and which is remembered by the activity of the neuron. 
We study the dynamical behaviors of this improved neuron model by 
changing external harmonic current and the magnetic gain parameters. The model shows rich dynamics including 
periodic/chaotic spiking and bursting, and remarkably, chaotic super-bursting, which has greater 
information encoding potentials than a standard bursting activity. 
Based on Krasovskii-Lyapunov stability theory, the sufficient conditions (on the synaptic strengths and magnetic gain parameters) 
for the chaotic synchronization of the improved model are obtained.
Based on Helmholtz’s theorem, the Hamilton function of the corresponding error dynamical system is also obtained.
It is shown that the time variation of this Hamilton function along trajectories can play the 
role of the time variation of the Lyapunov function -- in determining the asymptotic stability of the synchronization manifold.
Numerical computations indicate that as the synaptic strengths and the magnetic gain parameters change, 
the time variation of the Hamilton function is always non-zero (i.e., a relatively large positive or negative value) only when 
the time variation of the Lyapunov function is positive, 
and zero (or vanishingly small) only when the time variation of the Lyapunov function is also zero.
This clearly therefore paves an alternative way to determine the asymptotic stability of synchronization manifolds, and can be particularly 
useful for systems whose Lyapunov function is difficult to construct, but whose Hamilton function corresponding to the dynamic 
error system is easier to calculate.

\keywords{Neurons \and Magnetic flux \and Chaotic synchronization \and Lyapunov function \and Hamilton function }
\end{abstract}

%%%%%%%%%%%%%%%%%%%%%%%%%%%%%%%%%%%%%%%%%%%%%%%%%%%%%%%%%%%%%%%%%%%%%%%%%%%%%%%%%%%%%%%%%%%%%%%%%%%%%%%%%%%%%%%%%%%%%%%%%
\section{Introduction}\label{sect1}
%%%%%%%%%%%%%%%%%%%%%%%%%%%%%%%%%%%%%%%%%%%%%%%%%%%%%%%%%%%%%%%%%%%%%%%%%%%%%%%%%%%%%%%%%%%%%%%%%%%%%%%%%%%%%%%%%%%%%%%%%
\noindent
Mathematical modeling, dynamical systems theory, and numerical simulations are important tools for analyzing the complex dynamical behaviors 
of neural systems. The most biologically plausible  mathematical neuron model which describes the generation and 
transmission of action potential in neurons was proposed by Hodgkin and Huxley (HH model) in 1952 \cite{hodgkin1952quantitative}. 
Due to the very strong nonlinearity and slow computational speed of the HH model, 
mathematically simpler and computationally faster neuron models that still capture the qualitative behaviors of the HH model have been proposed. 
Some of the popular models includes: FitzHugh-Nagumo (FHN) \cite{fitzhugh1961impulses}, 
Hindmarsh-Rose \cite{hindmarsh1982model},  Morris-Lecar \cite{morris1981voltage}, and Izhikevich \cite{izhikevich2004model} models. 

The 2D Hindmarsh-Rose (HR) neuron model \cite{hindmarsh1982model} is more than ten times faster in computational speed \cite{izhikevich2004model} than the HH model. 
It is capable of producing some important behaviors such as spiking and sub-threshold oscillations -- which are also observed in real biological neurons -- upon variations of the model's parameters 
\cite{izhikevich2004model,wang2010lag,djati2011bidirectional}.
In order to capture other dynamical behaviors, such as chaos and bursting, observed in real biological neurons,
the original 2D HR neuron model has undergone few modifications. 
The main objective of Hindmarsh and Rose in 1984 (after the formulation of their 2-equations model \cite{hindmarsh1982model}) 
was to model the synchronization of firing of two snail neurons in
a simple way, without the use of the HH equations \cite{hindmarsh1982model,coombes2005bursting}. 
Hence, with the goal to design a neuron model that exhibits triggered firing, 
some modifications were done on the 2-equations model (by adding an adaptation variable, 
representing the slowly varying current,
that changed the applied current to an effective applied
one) to obtain the 3-equations model \cite{coombes2005bursting,hindmarsh1984model}.
This 3D model has been very popular in studying biological properties of
spiking and bursting neurons, including their chaotic dynamics.
Over the last decades, some detailed investigations and
studies of bifurcations and the dynamics of the 3D HR model has be done 
\cite{storace2008hindmarsh,gonzalez2003observation,gonzalez2007complex}, 
showing many behaviors observed of real biological neurons.

However, not only the 3D HR model fails to take into account the dynamics of calcium ions across the membrane,
it can only capture a relatively small domain of the chaotic regimes of real biological neuron.
A few years after the 3D HR model was proposed, Selverston et al. \cite{selverston2000reliable} 
studied a computational and electronic model of stomatogastric ganglion neurons. 
They used the standard 3D HR model, and discovered that in spite of the fact that 
the 3D model can produce several modes of spiking-bursting behaviors seen in biological neurons, 
its parameter space for chaotic activity is much more limited than the one observed in real neurons. 
For this reason, they modified the 3D HR model, by adding a fourth variable
(a slower process) representing the calcium dynamics \cite{selverston2000reliable}.
The complexity of the 3D  model increased and it was then capable of mimicking the complex dynamical 
(spiking, bursting and chaotic) behavior of pyloric central pattern generator neurons of the lobster 
stomatogastric system \cite{selverston2000reliable,de2008predicting}.
Thus, 4D HR model could capture more complex behavior than the 3D model 
\cite{selverston2000reliable,pinto2000synchronous} by simply taking into account the calcium exchange between 
intracellular warehouse and the cytoplasm, and in particular to completely produce the chaotic behavior of the stomatogastric ganglion neurons 
that the 3D model cannot.  Some detailed investigations and
studies of bifurcations and the dynamics of the 4D HR model has be done 
\cite{megam2014combined,ngouonkadi2016bifurcations,falcke2000modeling,rabinovich2002recovery}. 

Before the experimental confirmation of the non-negligible effects of the magnetic flux (generated by the flow of ions
across membrane) on the action potential of neurons, all previous studies on the dynamics of neuron models (including the
HH, FHN, Izhikevich, 3D and 4D HR models), have been done without taking into account magnetic flux effects.
Recently, M. Lv et al. \cite{lv2016model} proposed a modified HR model which takes into account the
effect of magnetic field by adding an additional variable for the magnetic flux
into the 3D HR model. This modified model not only can generate a variety of modes in electric activities by
changing the external forcing current and the magnetic flux parameters, but could also be useful in the investigation of the effect of electromagnetic fields 
on biological tissue \cite{wu2016model}.
All previous works on the memristive HR neuron model have considered only the 3D 
version of the HR model \cite{lv2016model,bao2018three,liu2019detecting,panahi2018complete,lv2016multiple,xu2018collective,usha2019hindmarsh},
leading to a 4D model including the memristive (magnetic flux) variable. 
Thus, for the first time, the memristive property of the 4D version of the HR model would be considered in this paper.
Including the memristive dynamics into the 4D HR model leads to a 5D HR model on which the study in this paper will focused on.
The aim of this paper is not to present a detailed bifurcation analysis of this improved 5D HR neuron model, but just to show the richer dynamical behavior the model can reproduce. 
On the other hand, because synchronization of memristive nonlinear systems has become an active area of research, and 
has widespread applications in secure communication and neuromorphic circuits \cite{abdurahman2015finite,abdurahman2015function}, 
the paper mainly focuses on the synchronization dynamics of the memristive 5D HR model. 

Using the memristive 5D HR neuron model,
an alternative way to determine the asymptotic stability of the synchronization manifold of coupled chaotic systems is presented.  
We show that the time variation of the Hamilton function of the error dynamical system associated to 
a pair of coupled chaotic 5D HR neurons can be used to determine the asymptotic stability of the synchronization
manifold, just as the time rate of change of Lyapunov function along trajectories of the system would do. 
It is shown that as the synaptic coupling strengths and memristive gain parameters of the model are varied,
we always have a non-zero time variation of this Hamilton function only when the time variation of the Lyapunov function
is positive (indicating an unstable synchronized state), and zero (or vanishingly small) only when the time variation of the Lyapunov function is also zero 
(indicating an asymptotically stable synchronized state). This indicates that the time variation of the Hamilton function associated to the error dynamical system of coupled oscillators
can be used as an asymptotic stability function. 

The rest of this paper is organized as follows: In Sect.\ref{sect2}, we present the improved 5D HR neuron model 
and show its rich dynamical behavior including spiking, bursting, super-bursting an chaos in time series, phase portraits and bifurcation diagrams.
In Sect.\ref{sect3}, we investigate the chaotic synchronization dynamics of a pair of neurons in the chaotic super-bursting regime 
and coupled via both time-delayed electrical and chemical synapses. The first part of this section is devoted to the 
analysis of the asymptotic stability of the synchronization manifold of the coupled neurons via the Lyapunov stability theory. 
The second part is devoted to the energy of the synchronization dynamics via the Hamilton function. 
Sect.\ref{sect4} deals with numerical simulations, showing the sign correlation between time variations of the Lyaponuv and the Hamilton function of the system.
In Sect.\ref{sect5}, we summary and concluding remarks.

%%%%%%%%%%%%%%%%%%%%%%%%%%%%%%%%%%%%%%%%%%%%%%%%%%%%%%%%%%%%%%%%%%%%%%%%%%%%%%%%%%%%%%%%%%%%%%%%%%%%%%%%%%
\section{Mathematical model and dynamics behavior}\label{sect2}
%%%%%%%%%%%%%%%%%%%%%%%%%%%%%%%%%%%%%%%%%%%%%%%%%%%%%%%%%%%%%%%%%%%%%%%%%%%%%%%%%%%%%%%%%%%%%%%%%%%%%%%%%%%
\noindent
The dynamical equation for the memristive 5D HR neuron model is described by
\begin{equation}\label{Eqn1}
\begin{split}
\left\{\begin{array}{lcl}
\displaystyle{ \frac{dx}{dt}}&=&-ax^3+bx^2+y-pz+I_0\cos(\varOmega t-\psi)\\
&-&k_1\rho(\phi) x,\\[1.0mm]
\displaystyle{\frac{dy}{dt}}&=&c-dx^2-y-\sigma w,\\[4.0mm]
\displaystyle{\frac{dz}{dt}}&=&r[s(x+x_0)-z],\\[4.0mm]
\displaystyle{\frac{dw}{dt}}&=&\mu[\gamma(y+y_0)-\delta w],\\[4.0mm]
\displaystyle{\frac{d\phi}{dt}}&=&x-k_2\phi,
\end{array}\right.
\end{split}
\end{equation}
where $x \in \mathbb{R}$ represents the membrane potential variable, $y\in \mathbb{R}$ is the recovery current variable associated to fast 
ions, $z\in \mathbb{R}$ is a slow 
adaption current associated to slow ions, $w\in \mathbb{R}$ represents an even slower process.
$I_0$ is the amplitude of a harmonic stimulus with frequency
$\varOmega$ and phase $\psi$. The other constant parameters have standard values: $a = 1.0$, $b = 3.0$, $p = 0.99$, $c = 1.01$, 
$d = 5.0128$, $\sigma = 0.0278$, 
$r = 0.00215$, $s = 3.966$, $x_0 = 1.605$, $\mu = 0.0009$, $\gamma = 3.0$, $y_0 = 1.619$, $\delta = 0.9573$.
The parameters $\mu<r\ll1$ play a
very important role in neuron activity; $r$ represents
the ratio of timescales between fast and slow fluxes across
the neuron’s membrane and $\mu$ controls the speed
of variation of the slower dynamical process $w$, in particular,
the calcium exchange between intracellular warehouse and
the cytoplasm \cite{selverston2000reliable,pinto2000synchronous}. 
The $\mu$ timescale induces richer dynamical behaviors including chaos in parameter regimes that the 
3D HR shows only periodic dynamics \cite{ngouonkadi2016bifurcations}. 

The fifth variable $\phi\in \mathbb{R}$, describes the 
magnetic flux across the neuron's cell membrane. The term $\rho(\phi) = \alpha + 3\beta\phi^2$ is the memory conductance of 
a magnetic-flux controlled
memristor \cite{muthuswamy2010implementing,bao2010steady,li2015hyperchaos,volos2015memristor}, 
where $\alpha$ and $\beta$ are fixed parameters which we will fixed throughout the paper at $\alpha=0.1$, $\beta=0.02$.
$k_1$ and $k_2$ are feedback gain; $k_1$ bridges the coupling and modulation on membrane potential $x$ 
from magnetic field $\phi$, and $k_2$ describes the degree
of polarization and magnetization by adjusting the saturation of magnetic flux \cite{ma2017mode}.
Following the Faraday's law of electromagnetic induction and the basic properties of a memristor, 
the term  $k_1\rho(\phi)v$ could be regarded as additive induction current on the membrane potential.
The dependence of electric charge $q$ on magnetic flux $\phi$ is defined by the memory-conductance as follows 
\cite{lv2016multiple,hong2013design}
\begin{equation}\label{eq:4hg}
\rho(\phi)=\frac{dq(\phi)}{d\phi} = \alpha + 3\beta\phi^2.
\end{equation}
Moreover, we know that current $i$ is defined by the time rate of charge $q$. Hence, 
the physical significance for the term $\rho(\phi)v$ could be described as follows
\begin{equation}\label{eq:5h} 
i=\frac{dq(\phi)}{dt}=\frac{dq(\phi)}{d\phi}\frac{d\phi}{dt}=\rho(\phi)V=k_1\rho(\phi)x,
\end{equation}
where  the  variable $V$ denotes the induced  electromotive  force,  which  holds  a  same  physical  unit, 
and parameter $k_1$ is the feedback gain. The ion currents of sodium and potassium contribute to the membrane potential and also 
the magnetic flux across the membrane; thus, a negative feedback term $-k_2\phi$ is introduced in the fifth equation of Eq.~\eqref{Eqn1}.

\begin{figure}[htp!]
\begin{center}
\includegraphics[width=0.25\textwidth]{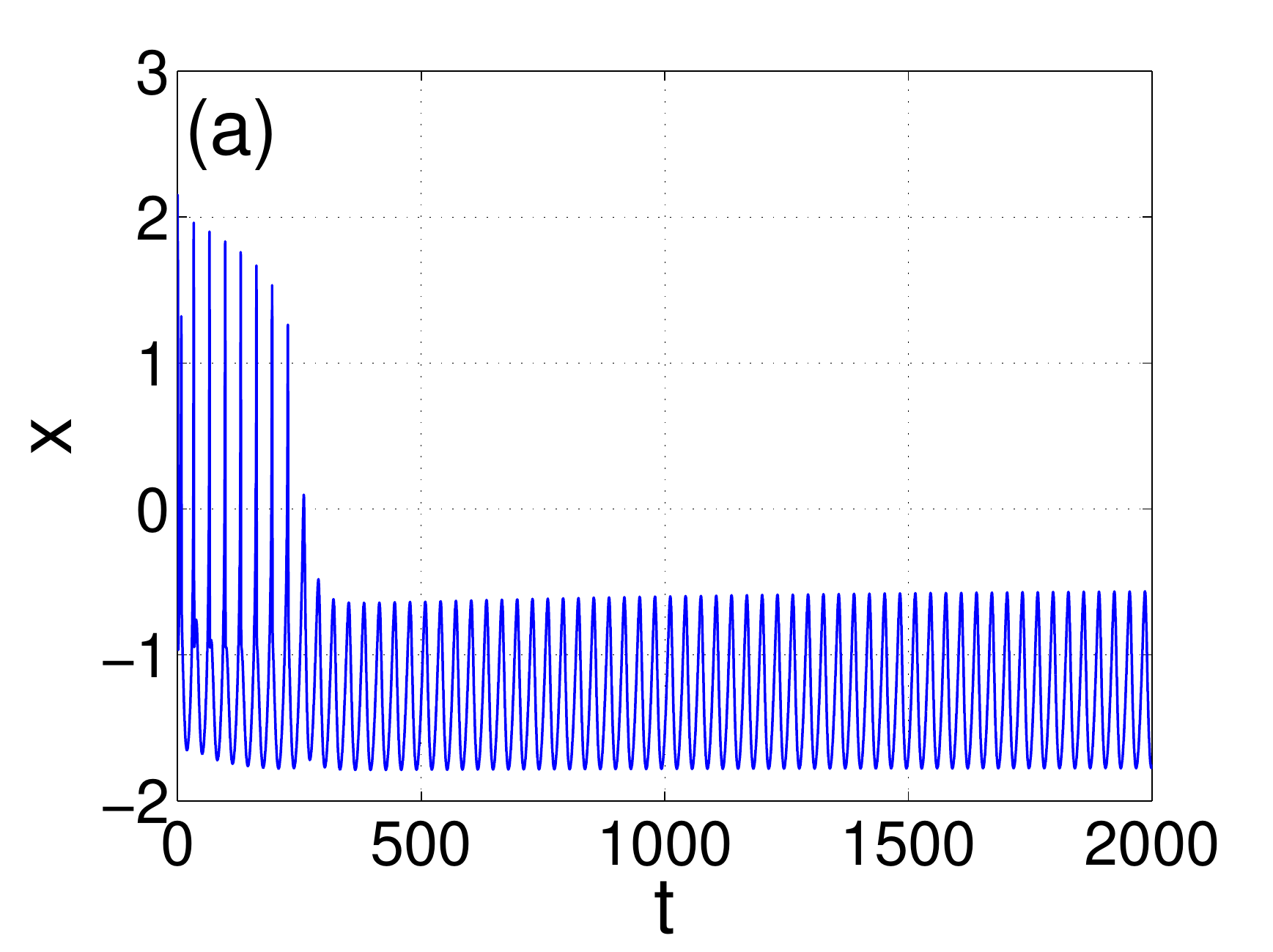}\includegraphics[width=0.25\textwidth]{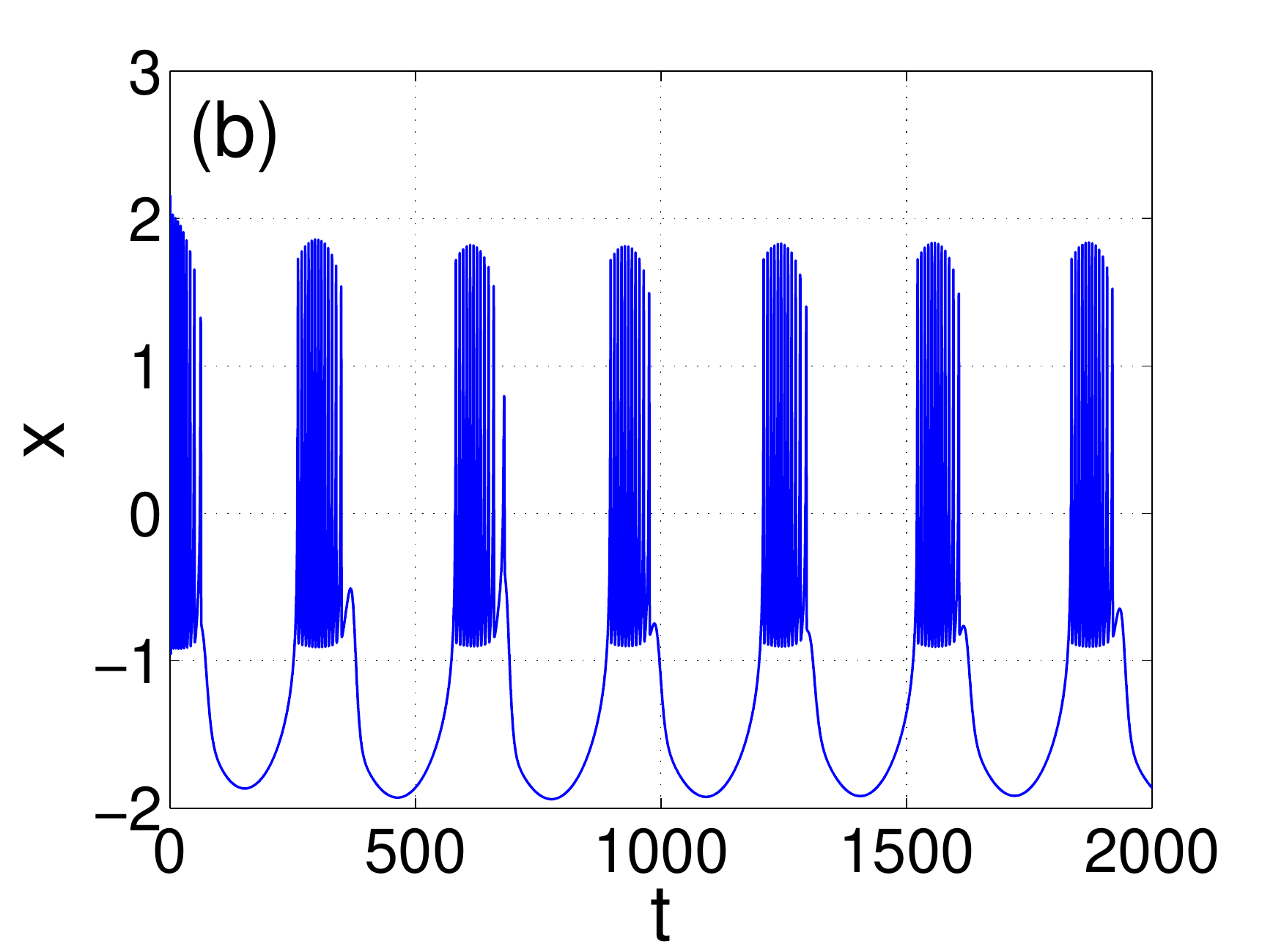}
\includegraphics[width=0.25\textwidth]{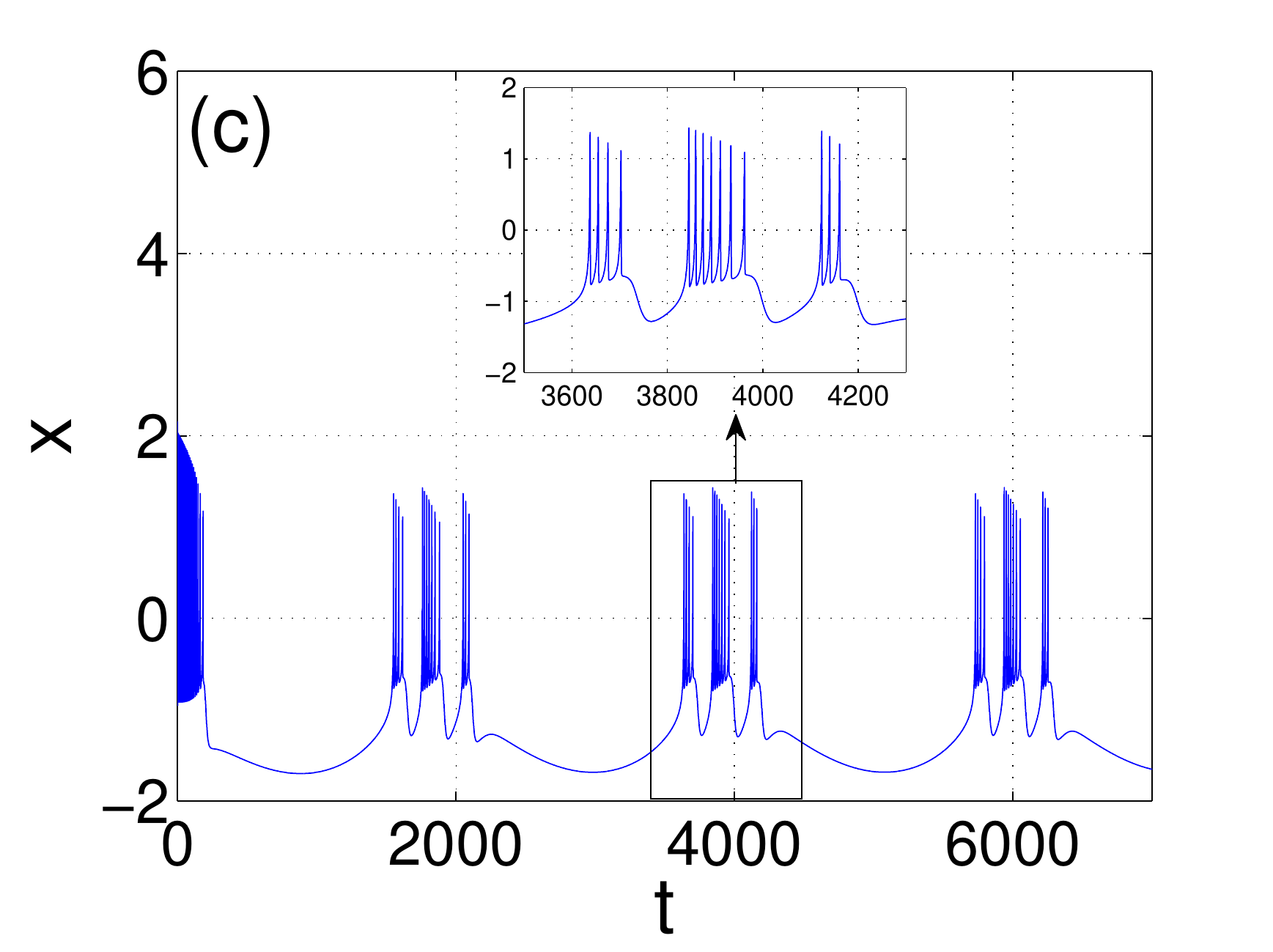}\includegraphics[width=0.25\textwidth]{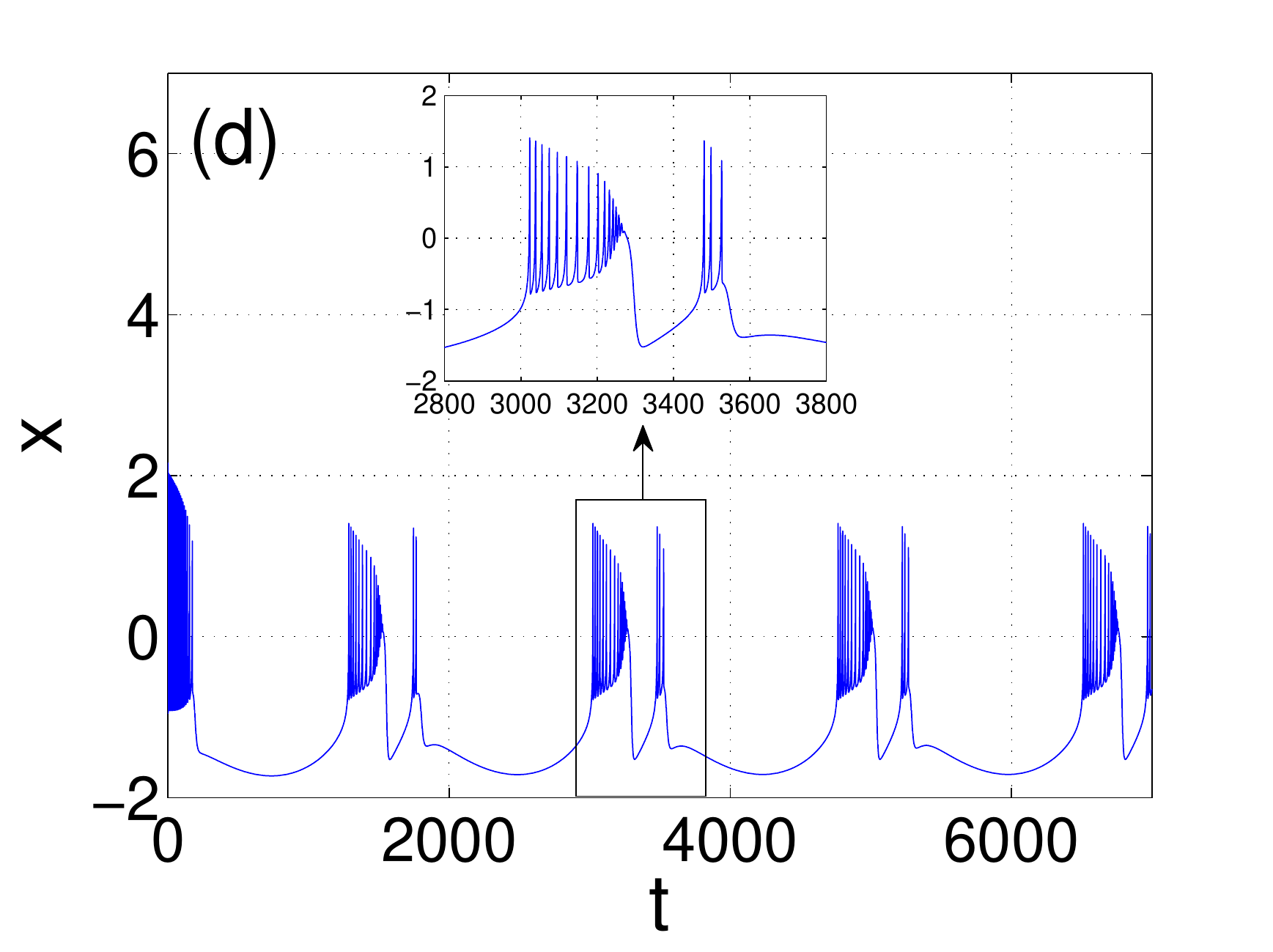}
\caption{The time series of the membrane potential $x$. 
\text{(a)}: $\Omega = 0.2$, periodic spiking activity. \text{(b)}: $\Omega = 0.02$, classical bursting activity. 
\text{(c)}: $\Omega = 0.003$, super-bursting activity consisting  of three bursts each. \text{(d)}: $\Omega = 0.0036$, 
super-bursting activity consisting  of only two bursts each. In all panels, the rest parameters are set at: 
$I_0 = 1.6$, $k_1 = 1.0$, $k_2 = 0.5$.
}\label{fig:1}
\end{center}
\end{figure}

We select the amplitude of harmonic forcing current $I_0$, its phase $\psi$, and the memristive gain parameters $k_1$ and $k_2$ as: $I_0 = 1.6$, $\psi=0.1$,
$k_1 = 1.0$, and $k_2 = 0.5$.  In Fig.\ref{fig:1}, the time series for  membrane potential $x$  under different values of the 
frequency $\Omega$  of the harmonic forcing are shown. The electrical activity of the 5D HR neuron show a rich dynamical behavior.
In Fig.\ref{fig:1}\text{(a)}, the model displays a simple periodic spiking activity with $\Omega=0.2$. In Fig.\ref{fig:1}\text{(b)}, 
$\Omega$ decreases to $0.02$ and the simple periodic spiking activity changes to the standard bursting activity, 
with each burst consisting of eleven or twelve spikes.
In Fig.\ref{fig:1}\text{(c)}, when the frequency is further reduced to $\Omega=0.003$, the standard bursting activity changes to 
a super-bursting activity, where each super-burst consist of three standard bursts which in turn consist of different number of spikes.
In Fig.\ref{fig:1}\text{(d)}, the frequency is increased by a little, i.e., to $\Omega=0.0036$, 
the pattern of the super-burst changes, with each super-burst consisting this time of only two standard bursts 
which have different spiking patterns.

The timing precision of the information processing in neural systems is very important
because the information in encoded at the spiking and bursting times \cite{pei1996noise}. 
This means that in our 5D neuron model in a super-bursting regime,
one kind of information could be encoded at super-burst time; then a second kind of information 
encoded at the standard burst times of each of these super-bursts; and then a third kind of information encoded 
at the spiking time of each standard burst of a super-burst. 
Whereas in a simple spiking regime, only one kind of information could be encoded at a time.
It is well known that bursts provide a more reliable mode of information transfer than spikes \cite{lisman1997bursts}.

In order to investigate the dependence of the system's behavior on its parameters -- the driving harmonic current ($I_0$, $\Omega$) 
and the memristive gain parameters ($k_1, k_2$) -- several bifurcation diagrams, each fully traced by its corresponding maximum Lyaponuv exponent, 
have been computed by using the peaks of the spikes in the super-bursting regime of Fig.\ref{fig:1}\text{(c)}. 
The diagrams shown have been chosen to illustrate the general dynamical structure of the memristive 5D HR neuron model. 
The maximum Lyaponuv exponent, $\Lambda_{max}$, of the model which is defined as
\begin{eqnarray}
\Lambda_{max}= \lim \limits_{\tau\rightarrow\infty}
\frac{1}{\tau}\ln\Big[\|L(\tau)\|\Big],
\end{eqnarray}
where $\|L(\tau)\|=\big(\delta x^2 + \delta y^2 + \delta w^2 +
\delta z^2+\delta\phi^2\big)^{1/2}$ is computed numerically by solving,
simultaneously, the system in Eq.~\eqref{Eqn1} and its corresponding
variational system of equations. Starting from an initial
condition, the system of Eq.~\eqref{Eqn1} is numerically integrated with the fourth-order Runge-Kutta algorithm and the
data recorded after some transient time.  A strictly negative maximum Lyapunov exponent characterizes an asymptotically stability system and the more negative the
exponent the greater the stability. When $\Lambda_{max}=0$, we
have a marginally stable system with quasi-periodic trajectories. And when $\Lambda_{max}>0$, the trajectories are unstable, 
meaning two nearby trajectories would diverge and the evolution of the system becomes sensitive to infinitesimal perturbations of initial conditions and hence chaotic \cite{kwuimy2008dynamics}.

\begin{figure}[htp!]
 \begin{center}
\includegraphics[width=0.25\textwidth]{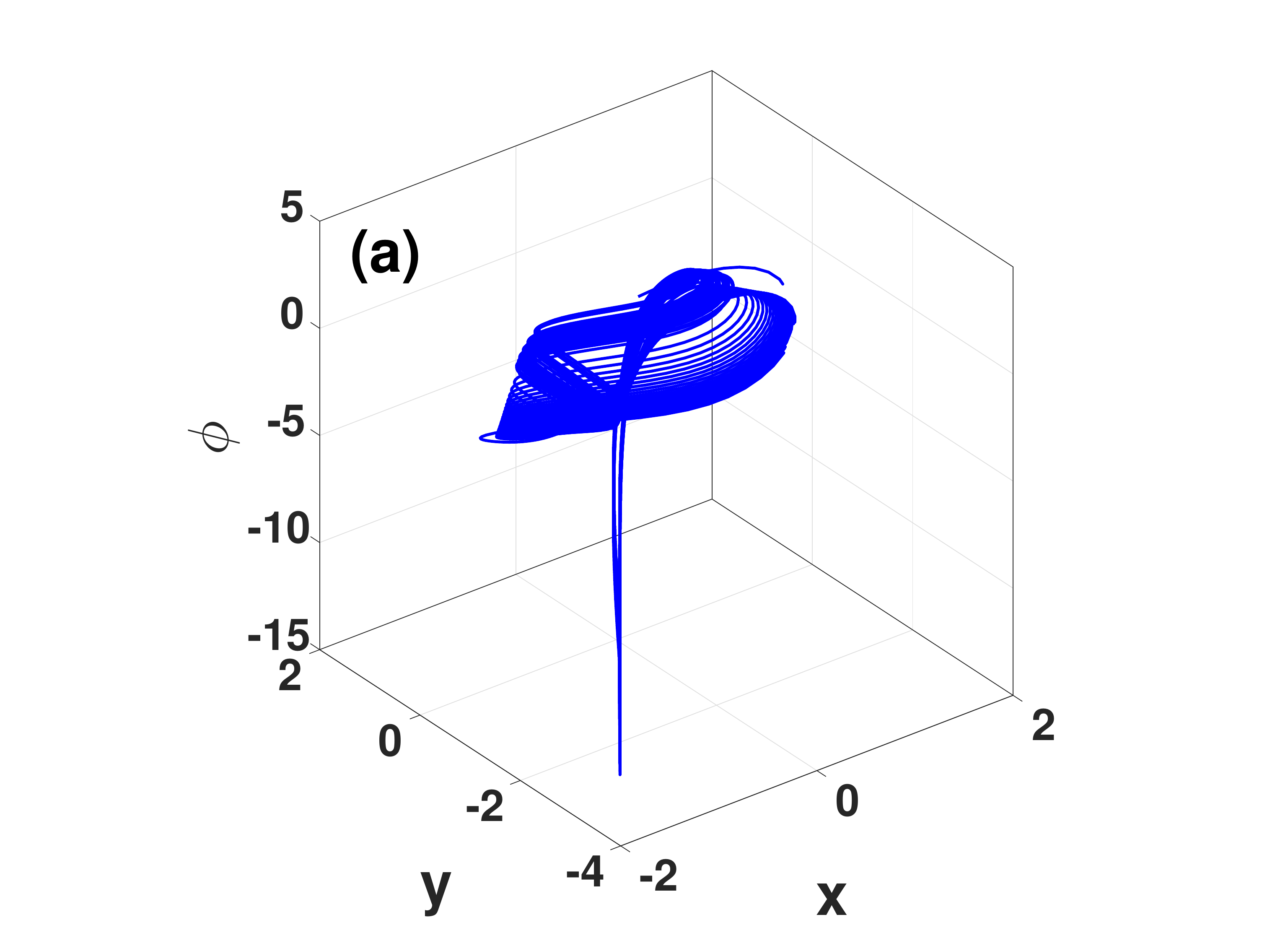}\includegraphics[width=0.25\textwidth]{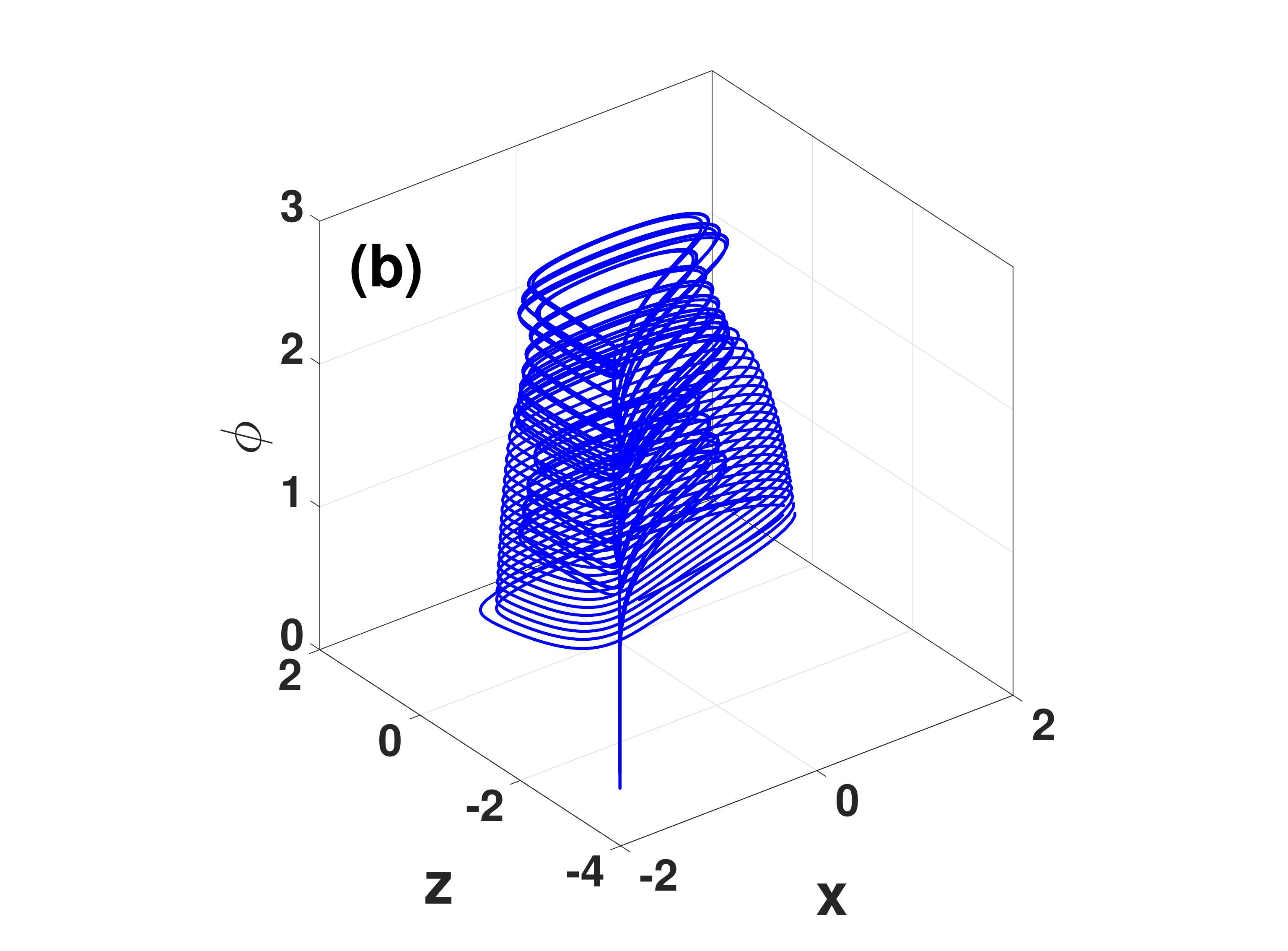}
\includegraphics[width=0.25\textwidth]{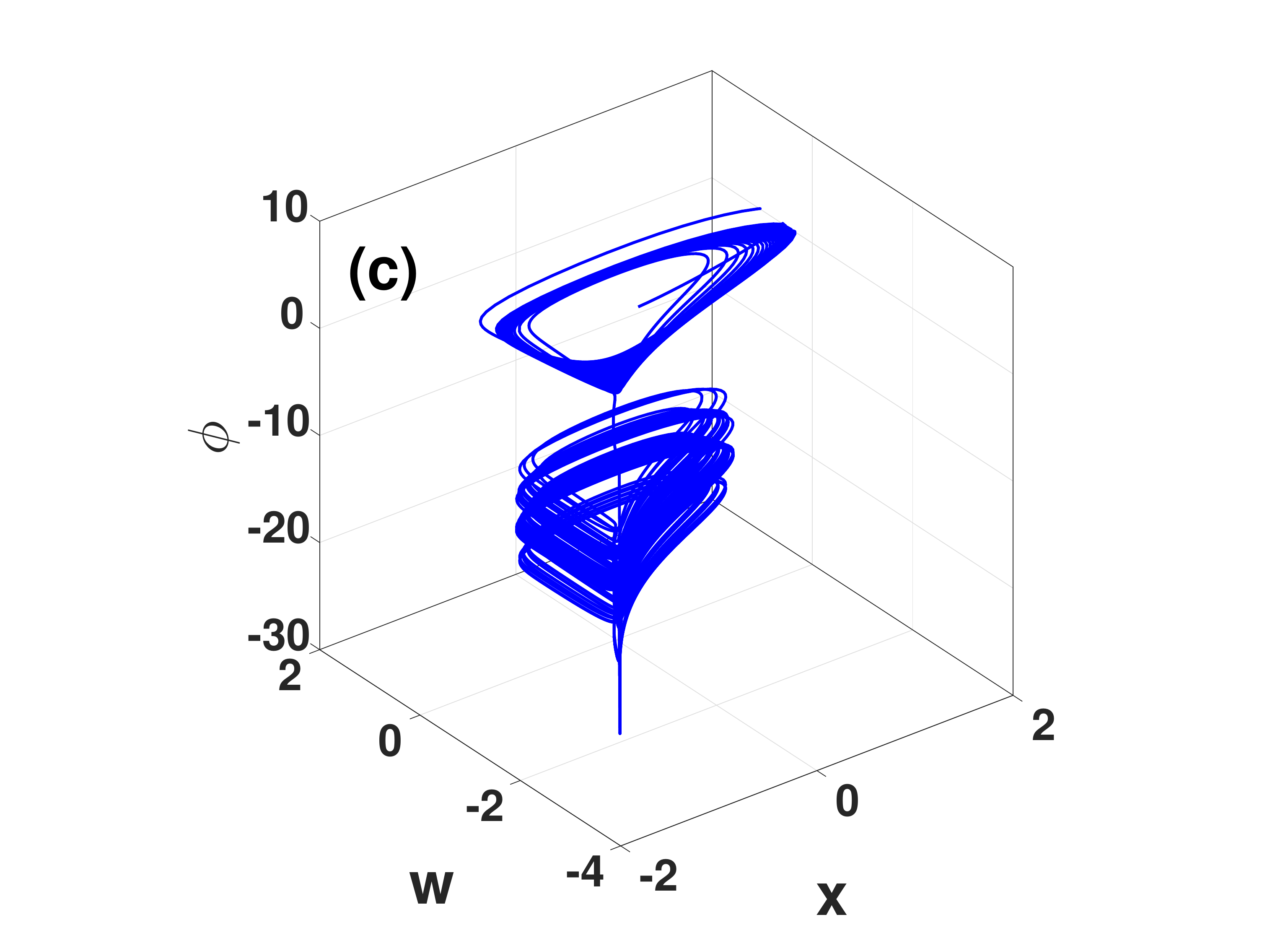}
 \caption{A bounded chaotic attractor of the improved 5D HR neuron in the super-bursting regime of Fig.\ref{fig:1}\text{(c)}, 
 projected onto 3 subspaces of the 5D phase space. \textbf{(a)} : $xy\phi$-space,  \textbf{(b)}: $xz\phi$-space,  \textbf{(c)}: $xw\phi$-space. The parameters are set at: 
 $I_0 = 1.6$, $\Omega = 0.003$, $k_1 = 1.0$, $k_2 = 0.5$.}\label{fig:2}
 \end{center}
\end{figure}

Fig.\ref{fig:2}\textbf{(a)} - \textbf{(c)} show the bounded phase space of a chaotic attractor of the 5D memristive HR neuron in a 
the super-bursting regime, projected onto the 3D subspaces of the 5D phase space: $xy\phi$-space, $xz\phi$-space, and $xw\phi$-space, 
respectively.

\begin{figure}[htp!]
\begin{center}
\includegraphics[width=0.52\textwidth]{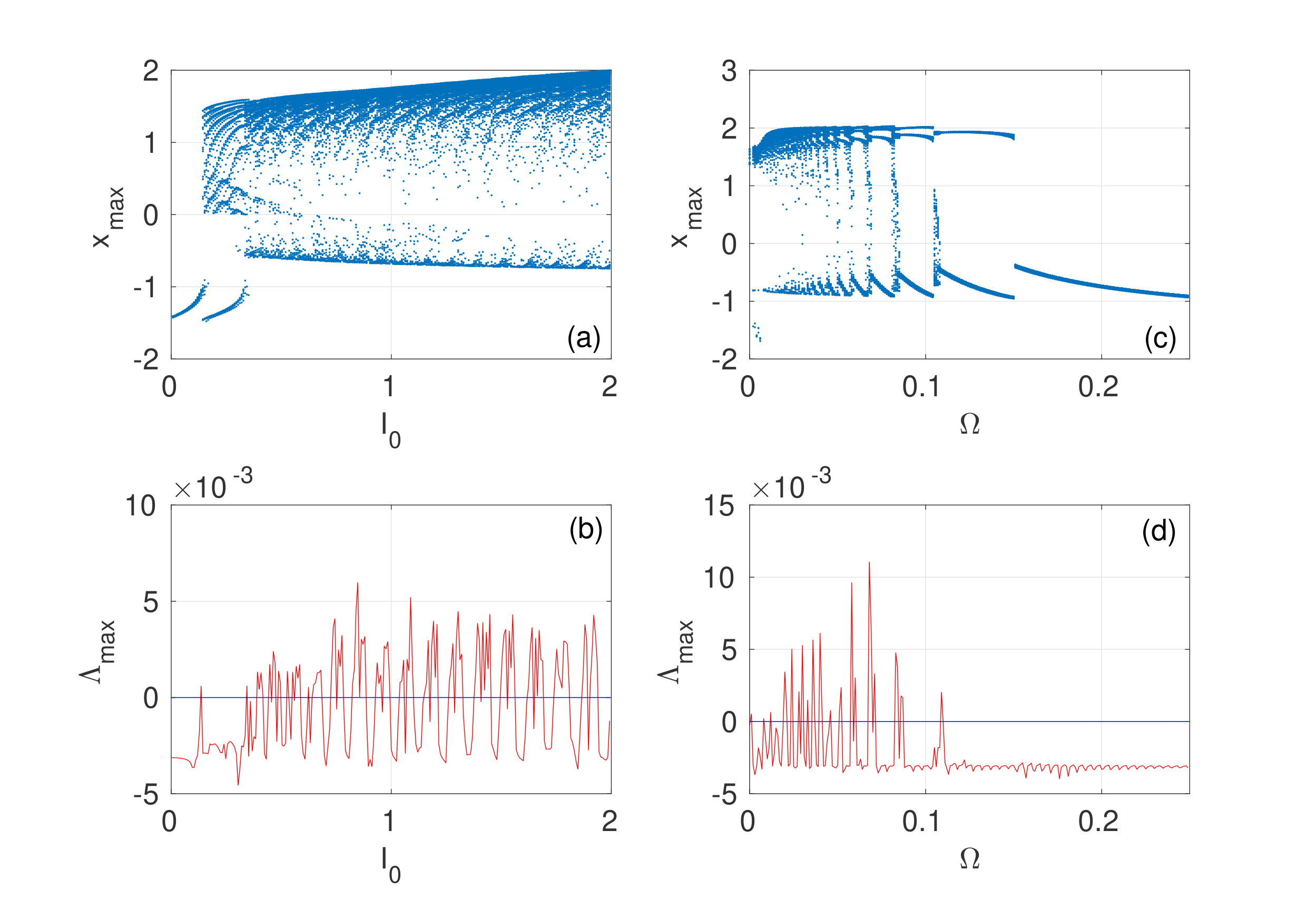}
\caption{Bifurcation diagrams each fully traced by its Lyapunov spectrum for the bifurcation parameters 
$I_0$ in \textbf{(a)}   and  \textbf{(b)} with $\Omega = 0.003$;
and $\Omega$ in \textbf{(c)}  and  \textbf{(d)} with $I_0= 1.6$. 
In all panels, $k_1 = 1.0$ and $k_2 = 1.2$.}\label{fig:3}
\end{center}
\end{figure}

In Fig.\ref{fig:3}\textbf{(a)} and \textbf{(b)}, we respectively show a bifurcation diagram
and its corresponding Lyapunov spectrum $\Lambda_{max}$. Different bifurcation sequences occur as amplitude of 
the external harmonic current is varied.
The system dynamics is mainly chaotic and it is intermingled with
thin windows of periodic orbits. 
In Fig.\ref{fig:3}\textbf{(c)} and \textbf{(d)},  different bifurcation sequences occur as of the frequency of 
the external harmonic current is varied.
The system dynamics is mainly chaotic for lower frequency values.

\begin{figure}[htp!]
\begin{center}
\includegraphics[width=0.52\textwidth]{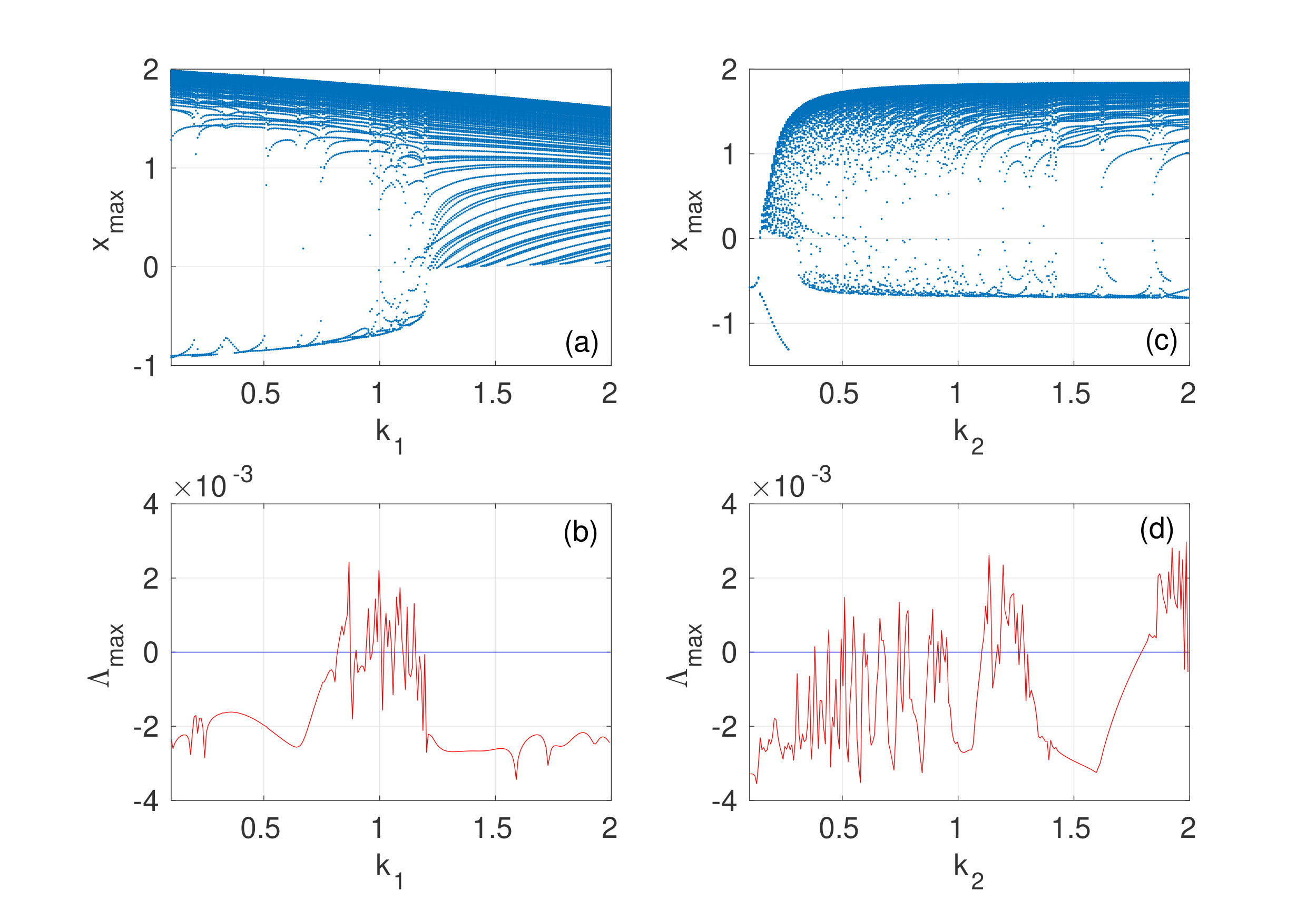}
\caption{Bifurcation diagrams each fully traced by
its Lyapunov spectrum for the bifurcation parameters $k_1$ in \textbf{(a)}  and \textbf{(b)} with fixed at $k_2 = 1.2$; 
and $k_2$ in \textbf{(c)} and \textbf{(d)} with fixed at $k_1= 1.0$. 
In all panels, $I_0 = 1.6$ and  $\Omega = 0.003$.}\label{fig:4}
\end{center}
\end{figure}

In Fig.\ref{fig:4}\textbf{(a)} and \textbf{(b)}, we respectively show a bifurcation diagram
and its corresponding Lyapunov spectrum $\Lambda_{max}$ for the memristive parameters $k_1$. 
The system dynamics is mainly chaotic for intermediate values of $k_1$. 
In Fig.\ref{fig:4}\textbf{(c)} and \textbf{(d)}, chaotic and periodic dynamics are intermingled as $k_2$ is varied.
In the next section of the paper, the values of the memristive parameters shall be fixed in a chaotic regime. 

%%%%%%%%%%%%%%%%%%%%%%%%%%%%%%%%%%%%%%%%%%%%%%%%%%%%%%%%%%%%%%%%%%%%%%%%%%%%%%%%%%%%%%%%%%%%%%%%%%%%%%%%%%
\section{Synchronization dynamics}\label{sect3}
%%%%%%%%%%%%%%%%%%%%%%%%%%%%%%%%%%%%%%%%%%%%%%%%%%%%%%%%%%%%%%%%%%%%%%%%%%%%%%%%%%%%%%%%%%%%%%%%%%%%%%%%%%%
This section deals with the main topic of this paper -- the chaotic synchronization dynamics of the coupled 5D HR neuron.
Here, we consider a pair HR neurons coupled via both time-delayed electrical and inhibitory 
chemical synapses. Electrical and chemical synapses are the two ways through which neurons connect 
to each other \cite{pereda2014electrical}. Electrical synapses
 connects the cytoplasm of nearby neurons directly and as a result the
 transmission of electrical impulses occur relatively quickly.
 The corresponding functional form of the bidirectional interaction mediated by the electrical synapses is
 defined as the difference between the membrane potentials of two adjacent neurons.
 On the other hand, with chemical synapse, the transmission of information takes place 
 via the release of a neurotransmitter. The functional form of this synaptic interaction 
 is considered as a nonlinear sigmoidal input-output function \cite{greengard2001neurobiology}.
 The system of coupled memristive 5D HR neurons is given as
\begin{equation}\label{Eqn7}
\begin{split}
\left\{\begin{array}{lcl}
\displaystyle{ \frac{dx_i}{dt}}&=&-ax_i^3+bx_i^2+y_i-pz_i+I_0\cos(\varOmega t-\psi)\\
&-&k_1(\alpha+3\beta\phi_i^2) x_i+g_{e}\xi_i(x_j-x_i)\\
&-&g_{c}(x_i-V_{syn})\sum\limits_{j=1}^{2}\eta_{ij}G(x_i,x_j),\\[1.0mm]
\displaystyle{ \frac{dy_i}{dt}}&=&c-dx_i^2-y_i-\sigma w_i,\\[4.0mm]
\displaystyle{ \frac{dz_i}{dt}}&=&r[s(x_i+x_0)-z_i],\\[4.0mm]
\displaystyle{ \frac{dw_i}{dt}}&=&\mu[\gamma(y_i+y_0)-\delta w_i],\\[4.0mm]
\displaystyle{ \frac{d\phi_i}{dt}}&=&x_i-k_2\phi_i,
\end{array}\right.
\end{split}
\end{equation}
with $i,j=\overline{1,2}$. The parameters $g_{e}$ and $g_{c}$ are the electrical and chemical coupling strengths, 
respectively. The chemical synaptic function is modeled by a sigmoidal nonlinear function $G(x_i,x_j)$ defined as:
$G(x_i,x_j):=G(x_j)=1/\big[1+\exp\{-\lambda(x_j-\theta_S)\}\big]$, where the parameter $\lambda = 10.0$ 
determines the slope of the function and $\theta_s = -0.25$ denotes the synaptic firing threshold.
 $V_{_\text{syn}}$ represents the synaptic reversal potential. For $V_{_\text{syn}} < x_{i}$ the chemical synaptic interaction has a 
depolarizing effect that makes the synapse inhibitory, and for $V_{_\text{syn}} > x_{i}$, the synaptic interaction has a 
hyper-polarizing effect making the synapse excitatory. 
For the 5D HR neuron model, the membrane potentials are bounded as 
$|x_{i}(t)| \leq 2.0$ ($i = 1,2$) for all times $t$. 
For the choice of fixed $V_{_\text{syn}} = -2.5$ (maintained throughout computations), 
the term ($x_{i}-V_{_\text{syn}}$) in Eq.~\eqref{Eqn7} is always positive. 
So, the inhibitory and excitatory natures of chemical synapses will depend only on the sign in front of the synaptic coupling strengths 
$g_{c}$. To make the chemical synapse inhibitory, we chose a negative sign.
The connectivity matrices $\xi_i$ and $\eta_{ij}$ are such that: $\xi_1=1$ and $\xi_2=-1$; 
$\eta_{ij}=\eta_{ji}=-1$ if $i\neq j$ and $\eta_{ij}=0$ if $i=j$. 
%%%%%%%%%%%%%%%%%%%%%%%%%%%%%%%%%%%%%%%%%%%%%%%%%%%%%%%%%%%%%%%%%%%%%%%%%%%%%%%%%%%%%%%%%%%%%%%%%%%%%%%%%%%%%%%%%%
\subsection{Asymptotic stability of synchronous states and Lyapunov function}
The complete synchronization of the coupled system in Eq.~\eqref{Eqn7} occurs when the two neurons asymptotically exhibit identical behavior, that is
\begin{equation}\label{Eqn8}
\begin{split}
\left\{\begin{array}{lcl}
\|x_2(t)-x_1(t)\|\rightarrow0,\\[0.1cm]
\|y_2(t)-y_1(t)\|\rightarrow0,\\[0.1cm]
\|z_2(t)-z_1(t)\|\rightarrow0,\\[0.1cm]
\|w_2(t)-w_1(t)\|\rightarrow0,\\[0.1cm]
\|\phi_2(t)-\phi_1(t)\|\rightarrow0,
\end{array}\right.
\end{split}
\end{equation}
as $t\rightarrow\infty$, for initial conditions chosen in some neighborhood of the synchronization manifold $\mathcal{M}_s$ on which
\begin{equation}\label{Eqn8a}
\begin{split}
\left\{\begin{array}{lcl}
x_1(t)=x_2(t)=x(t),\\[0.1cm]
y_1(t)=y_2(t)=y(t),\\[0.1cm]
z_1(t)=z_2(t)=z(t),\\[0.1cm]
w_1(t)=w_2(t)=w(t),\\[0.1cm]
\phi_1(t)=\phi_2(t)=\phi(t).
\end{array}\right.
\end{split}
\end{equation} 
The synchronization solution in Eq.~\eqref{Eqn8a} which satisfies the system of dynamical equations
\begin{equation}\label{Eqn8b}
\begin{split}
\left\{\begin{array}{lcl}
\displaystyle{ \frac{dx}{dt}}&=&-ax^3+bx^2+y-pz+I_0\cos(\varOmega t-\psi)\\
&-&k_1(\alpha+3\beta\phi^2) x-g_c(x-V_{syn})G(x),\\[1.0mm]
\displaystyle{\frac{dy}{dt}}&=&c-dx^2-y-\sigma w,\\[4.0mm]
\displaystyle{\frac{dz}{dt}}&=&r[s(x+x_0)-z],\\[4.0mm]
\displaystyle{\frac{dw}{dt}}&=&\mu[\gamma(y+y_0)-\delta w],\\[4.0mm]
\displaystyle{\frac{d\phi}{dt}}&=&x-k_2\phi.
\end{array}\right.
\end{split}
\end{equation}
where $x_j-x_i=0$, is also always a solution of coupled system in Eq.~\eqref{Eqn7}.
However, this synchronization solution might be stable only under some conditions.
By introducing coordinates transformation defined by directions transverse to the synchronization manifold $\mathcal{M}_s$ as 
\begin{equation}\label{Eqn10a}
\begin{split}
\left\{\begin{array}{lcl}
e_x=x_2-x_1,\\[0.1cm] 
e_y=y_2-y_1,\\[0.1cm] 
e_z=z_2-z_1,\\[0.1cm] 
e_w=w_2-w_1,\\[0.1cm] 
e_{\phi}=\phi_2-\phi_1,
\end{array}\right.
\end{split}
\end{equation}
we obtain the dynamics of the transverse perturbations to the synchronization manifold as
\begin{equation}\label{Eqn10}
\begin{split}
\left\{\begin{array}{lcl}
\displaystyle{\frac{de_x}{dt}}&=&-ae_x^3-2g_ee_x-\Bigg[3ax^2-2bx+k_1\alpha\\
&+& g_c\Big(1+e^{-\lambda(x-\theta_s)}\Big)^{-2}\Bigg]e_x+e_y-pe_z,\\[1.0mm]
\displaystyle{\frac{de_y}{dt}}&=&-2dxe_x-e_y-\sigma e_w,\\[4.0mm]
\displaystyle{\frac{de_z}{dt}}&=&r(se_x-e_z),\\[4.0mm]
\displaystyle{\frac{de_w}{dt}}&=&\mu(\gamma e_y-\delta e_w),\\[4.0mm]
\displaystyle{\frac{de_{\phi}}{dt}}&=&e_x-k_2e_{\phi}.
\end{array}\right.
\end{split}
\end{equation}
The following theorem can be obtained.
\begin{theorem}
  \textit{The coupled system in Eq.~\eqref{Eqn7} synchronizes if the five ordered-main sub-determinants in Appendix 
\eqref{eq:A1} are all strictly positive.}
\end{theorem}
\textit{{Proof:}} We construct a continuous, positive-definite Lyapunov function $V$ with continuous first partial derivative of the form
\begin{equation}
 V(e_x,e_y,e_z,e_w,e_{\phi})=\frac{1}{2}\Big[e_x^2 + e_y^2 + e_z^2 + e_w^2 + e_{\phi}^2\Big].
\end{equation}
The time derivative of the Lyapunov function $V$ along trajectories of the error dynamical system in Eq.~\eqref{Eqn10} yields
\begin{eqnarray}\label{eqn:v}\nonumber
\frac{dV}{dt}=&-&ae_x^4-2g_ee_x^2-e_y^2-re_z^2-\mu\delta e_w^2-k_2e_{\phi}^2\\\nonumber
&+&(1-2dx)e_xe_y+(rs-p)e_xe_z+e_xe_{\phi} \\\nonumber
&+& (\mu\gamma-\sigma)e_ye_w-\Bigg[3ax^2-2bx+k_1\alpha\\
&+& g_c\Big(1+e^{-\lambda(x-\theta_s)}\Big)^{-2}\Bigg]e_x^2.
\end{eqnarray}
Since chaotic systems have bounded phase space, there exists a positive constant $J$, such that $\left|x(t)\right|<J$, thus
\begin{eqnarray}\label{eqn11}\nonumber
\frac{dV}{dt}\leq&-&ae_x^4-2g_e\left|e_x\right|^2-e_y^2-re_z^2-\mu\delta e_w^2\\\nonumber
&-&k_2e_{\phi}^2+(1-2dJ)\left|e_x\right|\left|e_y\right|+\left|e_x\right|\left|e_{\phi}\right|\\\nonumber
&+&(rs-p)\left|e_x\right|\left|e_z\right|+(\mu\gamma-\sigma)\left|e_y\right|\left|e_w\right|\\\nonumber
&-&\Bigg[3aJ^2-2bJ+k_1\alpha+g_c\Big(1+e^{-\lambda(J-\theta_s)}\Big)^{-2}\Bigg]e_x^2,
\end{eqnarray}
for every points of the attractor and can be compactly written as
\begin{eqnarray}\label{eqn11a}
\frac{dV}{dt}\leq-ae_x^4-2g_e\left|e_x\right|^2-E^T\textbf{M}E, 
\end{eqnarray}
where $E^T=\Big(e_x\:\:\:e_y\:\:\:e_z\:\:\:e_w\:\:\:e_{\phi}\Big)$ is a row vector with transpose $E$, and the matrix \textbf{M} is given by
\begin{eqnarray}\label{eq:1aa}
\textbf{M}=\left[\begin{array}{ccccc}
\mathbf{D_1}\:& (2dJ-1) \: & (p-rs) \:& 0 \:& -1	\\\\
(2dJ-1)\: & 1 \:& 0 \:& (\sigma-\mu\gamma) \:& 0	\\\\
(p-rs) \:& 0 \:& r \:& 0 \:& 0	\\\\
0 \:& (\sigma-\mu\gamma) \:& 0 \:& \mu\delta \:& 0   \\\\
-1\: & 0 \:& 0 \:& 0 \:& k_2
\end{array}\right],
\end{eqnarray}
with $\displaystyle{\mathbf{D_1}=(3aJ-2b)J+k_1\alpha+g_c\Big(1+e^{-\lambda(J-\theta_s)}\Big)^{-2}}$.

To ensure that the origin of the error dynamical system in Eq.~\eqref{Eqn10} is asymptotically stable, three conditions have to be met 
in Eq.~\eqref{eqn11a}: $a\geq0$ (which is always the case since $a=1.0$ 
in the model), $g_e\geq0$, and most importantly, the matrix \textbf{M} must be positive definite. \textbf{M} is positive definite if it 
satisfies \\Sylvester’s criterion given in Appendix \ref{eq:A2} 
-- in which case $dV/dt$ in Eq.~\eqref{eqn11a} will be negative semi-definite. 
According to the Lyapunov stability theory \cite{krasovskii1963stability} and Barbalat’s lemma \cite{popov1973hyperstability}, 
all the transverse perturbations decay to the synchronization manifold without any transient growth, i.e., one obtains
\begin{equation}\label{Eqn10aa}
\begin{split}
\left\{\begin{array}{lcl}
e_x(t)\rightarrow0,\\[0.1cm] 
e_y(t)\rightarrow0,\\[0.1cm] 
e_z(t)\rightarrow0,\\[0.1cm] 
e_w(t)\rightarrow0,\\[0.1cm] 
e_{\phi}(t)\rightarrow0,
\end{array}\right.
\end{split}
\end{equation}
as $t\rightarrow\infty$. It follows that the coupled system in Eq.~\eqref{Eqn7} synchronizes when the inequalities 
in Eq.~\eqref{eq:A2} are satisfied. This completes the proof. $\Box$

\subsection{Synchronization energy and Hamilton function}
Here, we determine the Hamilton energy function\\$H(e_x,e_y,e_z,e_w,e_{\phi})$ associated to the error system in Eq.~\eqref{Eqn10}. 
Using this energy function, we analytically evaluate the energy variation of the
coupled system. This energy variation is an important measure because it gives us the flow of energy 
in the process of synchronization and hence, it can be considered to be the amount of
energy per unit time needed to maintain a particular degree of synchrony \cite{torrealdea2006energy}. 
We focus on the effects of the synaptic couplings strengths $(g_e,g_c)$and magnetic flux parameters $(k_1,k_2)$ 
on this energy variation and compare it to the time variation of the Lyapunov function previously calculated.

Based on Helmholtz's theorem \cite{kobe1986helmholtz}, we express the velocity vector field $F(e_x,e_y,e_z,e_w,e_{\phi})$ of
the error dynamical system in Eq.~\eqref{Eqn10} as the sum of two vector fields: a divergence-free vector and a gradient vector field, i,e.,
$F(e_x,e_y,e_z,e_w,e_{\phi}) = f_c(e_x,e_y,e_z,e_w,e_{\phi}) \\+ f_d(e_x,e_y,e_z,e_w,e_{\phi})$, with  the 
conservative part\\$f_c(e_x,e_y,e_z,e_w,e_{\phi})$, containing the full rotation 
and the dissipative part, $f_d(e_x,e_y,e_z,e_w,e_{\phi})$, containing the whole divergence. A divergence-free vector 
and a gradient vector field of the error dynamical system in Eq.~\eqref{Eqn10} are given by
\begin{eqnarray}\label{Eqn:1bb}
f_c &=& \left(\begin{array}{c}
e_y-pe_z\\
-2dxe_x-\sigma e_w\\
rse_x\\
\mu\gamma e_y\\
e_x
\end{array}\right),\\
f_d &=& \left(\begin{array}{c}
-ae_x^3-2g_ee_x-\mathbf{N}e_x\\
-e_y\\
-re_z\\
-\mu\delta e_w\\
-k_2e_{\phi}
\end{array}\right),
\end{eqnarray}
where $\mathbf{N}=3ax^2-2bx+k_1\alpha+g_c\big(1+e^{-\lambda(x-\theta_s)}\big)^{-2}$.

For the conservative vector field, the equation
\begin{eqnarray}
\nabla H^Tf_c(e_x,e_y,e_z,e_w,e_{\phi})=0,
\end{eqnarray}
where $\nabla H$ denotes the transpose gradient of function $H$, defines a partial 
differential equation from which a Hamilton energy function (a generalized Hamiltonian) $H(e_x,e_y,e_z,e_w,e_{\phi})$ can be evaluated 
(see also \cite{chun2016calculation} for a detailed and general description of the method).
Thus, Hamilton energy function of the error dynamical system in Eq.~\eqref{Eqn10} satisfies the partial differential equation given by 
\begin{align}\label{1cc}\nonumber
&(e_y-pe_z)\frac{\partial H}{\partial e_x}-(2dxe_x+\sigma e_w)\frac{\partial H}{\partial e_y}\\
&+rse_x\frac{\partial H}{\partial e_z}+\mu\gamma e_y\frac{\partial H}{\partial e_w}+e_x\frac{\partial H}{\partial e_{\phi}}=0.
\end{align}
By the method of separation of variables, one solution of Eq.~\eqref{1cc}  is given by  
\begin{eqnarray}\label{1dd}\nonumber
H(e_x,e_y,e_z,e_w,e_{\phi})&=&Q\Bigg[(2dx+rsp-\mu\gamma\sigma)e_x^2+e_y^2\\\nonumber
&+&2\sigma e_xe_w+p\Big(p-\frac{\mu\gamma\sigma}{rs}\Big)e_z^2\\
&+&2p(1-e_y)e_z-2rspe_{\phi}\Bigg],
\end{eqnarray}
where $Q$ is an arbitrary constant.

Since any positive definite quadratic form can always be a solution for the energy partial differential equation 
compatible with the generalized Hamiltonian formalism, independently of the system at hand, the same trivial positive definite quadratic form can
always be assigned to different chaotic systems. However, assigning the same form of energy function to all chaotic systems 
fails to reveal the individual features of its own dynamics.
Hence, the approach used to obtain the Hamilton energy function $H$ requires
additional hypothesis in order to be able to assign to the error dynamical system a specific energy function. 
This additional hypothesis is set by introducing a connection between the
the change in the volume of the phase space of the coupled system and time rate of change of energy, as one cannot occur without the other. 
Hence, since the energy function in Eq.~\eqref{1dd} is not unique for the system, we compute
the time rate of change of this energy function along trajectories of the error dynamical system. 
This time rate of change of the Hamilton energy function is related to the divergence of the vector field, $f_d(e_x,e_y,e_z,e_w,e_{\phi})$,
responsible for contraction of the volume of the phase space. 
Thus, the time variation of the Hamilton energy function $H(e_x,e_y,e_z,e_w,e_{\phi})$ associated to the error dynamical
system in Eq.~\eqref{Eqn10} now becomes uniquely related to the specific dynamics of the error dynamical system.
This energy is dissipated via the dissipative component of the velocity vector field $f_d(e_x,e_y,e_z,e_w,e_{\phi})$ according to the equation \cite{chun2016calculation}
\begin{equation}\label{1ee}
\begin{split}
\frac{dH}{dt}=\nabla H^Tf_d.
\end{split}
\end{equation}
From Eq.~\eqref{Eqn:1bb}, Eq.~\eqref{1dd} and Eq.~\eqref{1ee}, we obtain the time variation of the Hamilton energy function as
\begin{eqnarray}\label{1ff}\nonumber
\frac{dH}{dt}&=&2Q\Bigg[rspk_2e_{\phi}+pe_ye_z+e_y^2+p\mu\delta e_ye_w-a\sigma e_x^3e_w\\\nonumber
&-&p\mu\delta e_w-\sigma g_ee_xe_w-p\mu\delta\Big(p-\frac{\mu\gamma\sigma}{rs}\Big)e_ze_w\\\nonumber
&-&\sigma\Big[3ax^2-2bx+k_1\alpha+g_c\big(1+e^{-\lambda(x-\theta_s)}\big)^{-2}\\\nonumber
&+&\mu\delta\Big]e_xe_w-\Big[ae_x^4+2g_ee_x^2+3ax^2e_x^2-2bxe_x^2\\\nonumber
&+&k_1\alpha e_x^2+g_c\big(1+e^{-\lambda(x-\theta_s)}\big)^{-2}e_x^2\Big]\\
&\times&\Big(2dx+rsp-\mu\gamma\sigma\Big)\Bigg].
\end{eqnarray}
Irrespective of the value of the arbitrary constant $Q$ (which we will fixed at $Q=-1.0$ throughout the rest of the paper), 
Eq.~\eqref{1ff} gives us an energy function which is unique to the 
error dynamical system of our coupled neurons, and it gives us information 
about the energy dissipated during the synchronization dynamics.
In the next section, we show (and provide a theoretical explanation) that the time variation of the Hamilton function of the error dynamical 
associated to the coupled neuron system paves an alternative way of determining the asymptotic stability of the 
synchronized state of the system just as the time variation of the Lyapunov function would do.

% %%%%%%%%%%%%%%%%%%%%%%%%%%%%%%%%%%%%%%%%%%%%%%%%%%%%%%%%%%%%%%%%%%%%%%%%%%%%%%%%%%%%%%%%%%%%%%%%%%%%%%%%%%
\section{Numerical simulations and discussion}\label{sect4}
% %%%%%%%%%%%%%%%%%%%%%%%%%%%%%%%%%%%%%%%%%%%%%%%%%%%%%%%%%%%%%%%%%%%%%%%%%%%%%%%%%%%%%%%%%%%%%%%%%%%%%%%%%%%
In this section, we compare the derivatives of the Lyaponuv function and the Hamilton energy function 
given in Eq.~\eqref{eqn:v} and Eq.~\eqref{1ff}, respectively, as the synaptic strengths ($g_e$ and $g_c$) and the memristive gain parameters 
($k_1$ and $k_2$) vary. To simulate these equations, we simultaneously integrate Eq.~\eqref{Eqn8b} and Eq.~\eqref{Eqn10} for a very large 
time interval using the fourth-order Runge Kutta algorithm. After discarding the transient time, 
we compute the mean values of $dV/dt$ and $dH/dt$.

\begin{figure}[htp!]
\begin{center}
\includegraphics[width=0.25\textwidth]{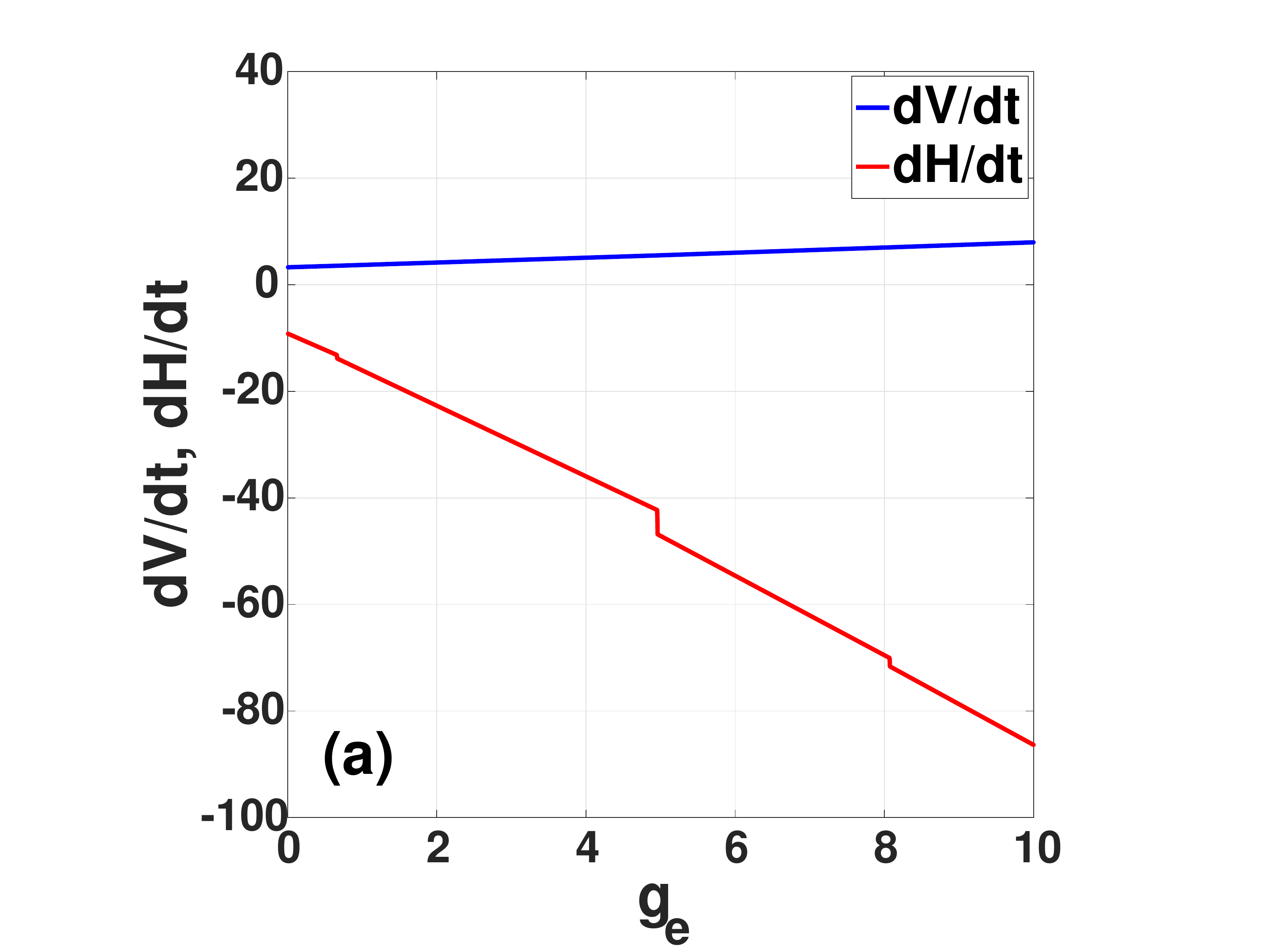}\includegraphics[width=0.25\textwidth]{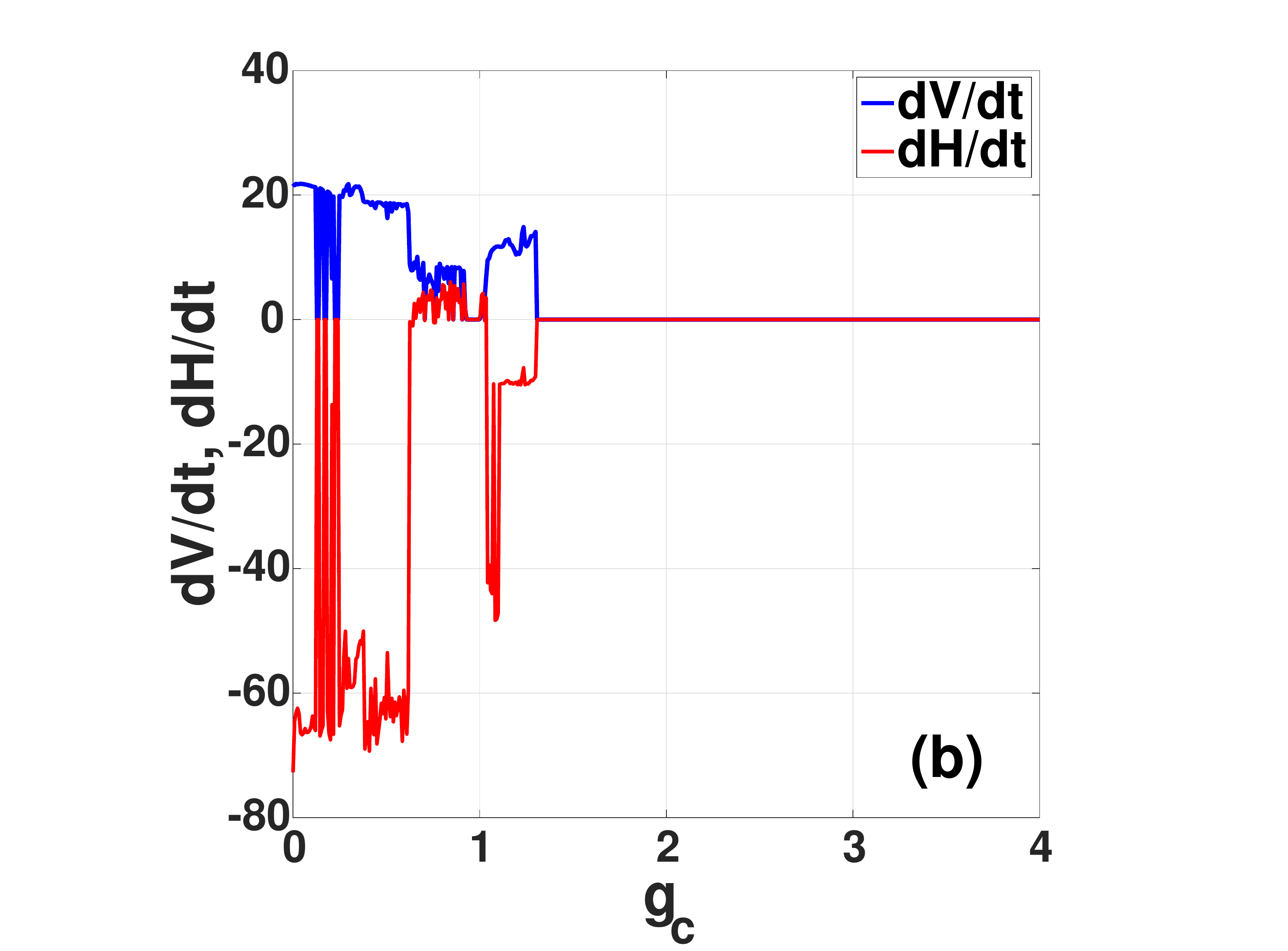}
\caption{Variations of $dV/dt$ and $dH/dt$ 
respectively represented by the blue and red curves, with respect to the synaptic strengths $g_e$ an $g_c$ 
for fixed memristive gain parameter values. $dV/dt>0$, indicating an unstable synchronization manifold $\mathcal{M}_s$ 
only $dH/dt<0$. We have $dV/dt=0$, for an asymptotically stable $\mathcal{M}_s$, only when $dH/dt=0$.
In panel \textbf{(a)}: $k_1=1.0, k_2=0.5, g_c=1.0$. In panel \textbf{(b)}: 
$k_1=1.0, k_2=0.5, g_e=1.5$.}\label{fig:5}
\end{center}
\end{figure}

For a weak chemical synaptic strength $g_c$, Fig.\ref{fig:5}\textbf{(a)} shows the variations of $dV/dt$ and 
$dH/dt$ with respect to electrical synaptic strength $g_e$.  It is observed 
that synchronization manifold $\mathcal{M}_s$ is always unstable as indicated by $dV/dt>0$ (blue curve). 
While for these same values of $g_e$, we always have $dH/dt<0$ (red curve).

In Fig.\ref{fig:5}\textbf{(b)}, we show the variations of $dV/dt$ and $dH/dt$ with respect to the chemical synaptic strength 
$g_c$ for a weak electrical synaptic strength $g_e$. Here, we see that $dH/dt<0$ only when $\mathcal{M}_s$ is unstable as indicated 
by $dV/dt>0$; and $dH/dt=0$ only when $\mathcal{M}_s$ is stable as indicated by $dV/dt=0$. 

\begin{figure}[htp!]
\begin{center}
\includegraphics[width=0.25\textwidth]{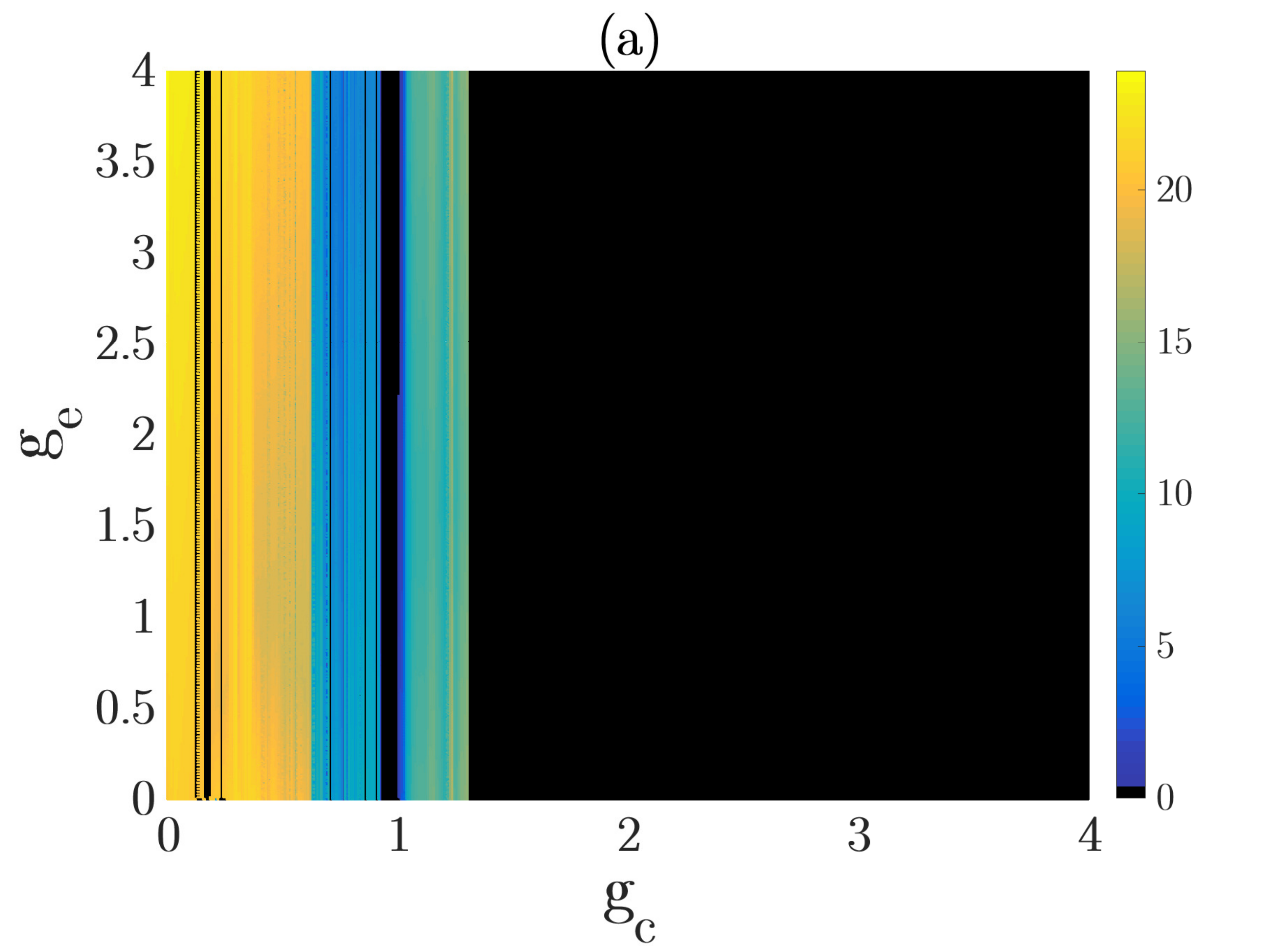}\includegraphics[width=0.25\textwidth]{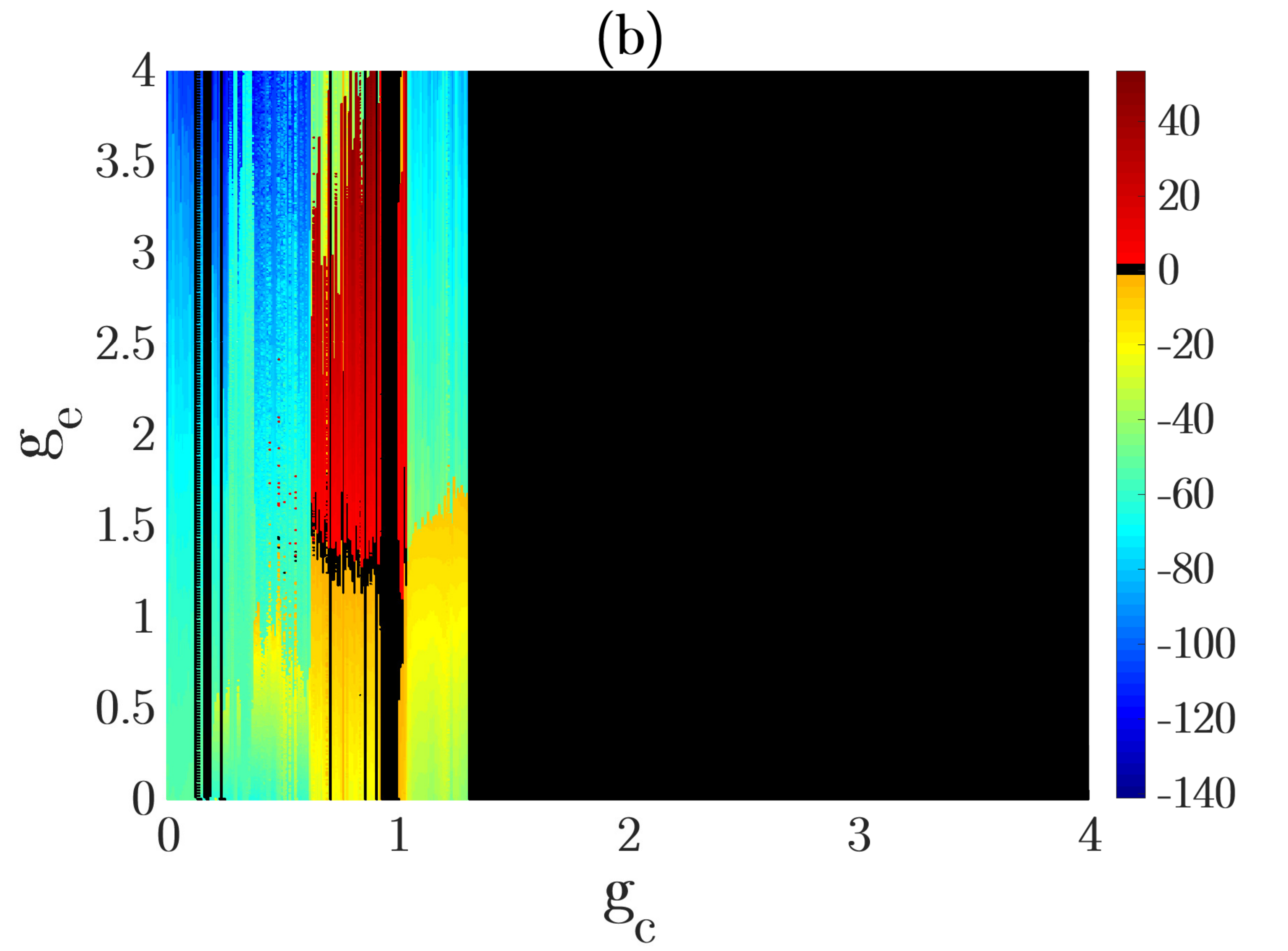}
\caption{Color-coded variations of $dV/dt$ in \textbf{(a)}) and $dH/dt$ 
in \textbf{(b)}) with respect to the synaptic coupling strengths for fixed memristive gain parameter values. 
The synchronization manifold $\mathcal{M}_s$ is unstable when $dV/dt>0$ which occurs only when $0<dH/dt<0$; 
and stable when $dV/dt=0$ which occurs only when $dH/dt=0$ or $dH/dt\approx0$ (i.e., vanishingly small).
Other parameters: $k_1=1.0, k_2=0.5.$}\label{fig:6}
\end{center}
\end{figure}

In Fig.\ref{fig:6}\textbf{(a)} and \textbf{(b)}, we show a color-coded global behavior of the $dV/dt$ and $dH/dt$
with respect to the synaptic coupling strength parameters $g_e$ and $g_c$ for fixed memristive gain parameters. Here we oberve 
that the sign correlation between  $dV/dt$ and $dH/dt$ persist for all values of the parameters, and not just for particular values.
Here, we have $0<dH/dt<0$ only when $dV/dt>0$, indicating an unstable synchronization manifold. 
And $dH/dt=0$ (or vanishingly small, $dH/dt\approx0$) only when $dV/dt=0$, indicating a stable synchronization manifold. 

In Fig.\ref{fig:6}\textbf{(a)}, the color-bar shows the variation of 
$dV/dt$ for the given range of values of $g_e$ and $g_c$ and indicates that for a sufficiently strong chemical coupling 
strength, i.e., $g_c>1.25$, the synchronization manifold becomes and stays stable, (i.e., $dV/dt=0$)
irrespective of the electrical synaptic strength $g_e$. 

In Fig.\ref{fig:6}\textbf{(b)}, where the color-bar shows the variation of $dH/dt$, we 
observed that $dH/dt=0$, for $g_c>1.25$ irrespective of the value of $g_e$ 
just as with $dV/dt$ in Fig.\ref{fig:6}\textbf{(a)}. 
For weak chemical coupling, i.e., $g_c<1.25$, we have $0<dH/dt<0$ (in Fig.\ref{fig:6}\textbf{(b)}) for the same parameter values of 
$g_e$ and $g_c$ for which $dV/dt>0$ (in Fig.\ref{fig:6}\textbf{(a)}), indicating unstable synchronized dynamics. 
However, in this weak chemical coupling regime ($g_c<1.25$), we can have some values of $g_c$ 
(e.g., $g_c=0.25$ and $g_c=1.0$) for which we have stable synchronized dynamics, i.e., $dV/dt=0$ with $dH/dt=0$ or $dH/dt\approx0$.   

\begin{figure}[htp!]
\begin{center}
\includegraphics[width=0.25\textwidth]{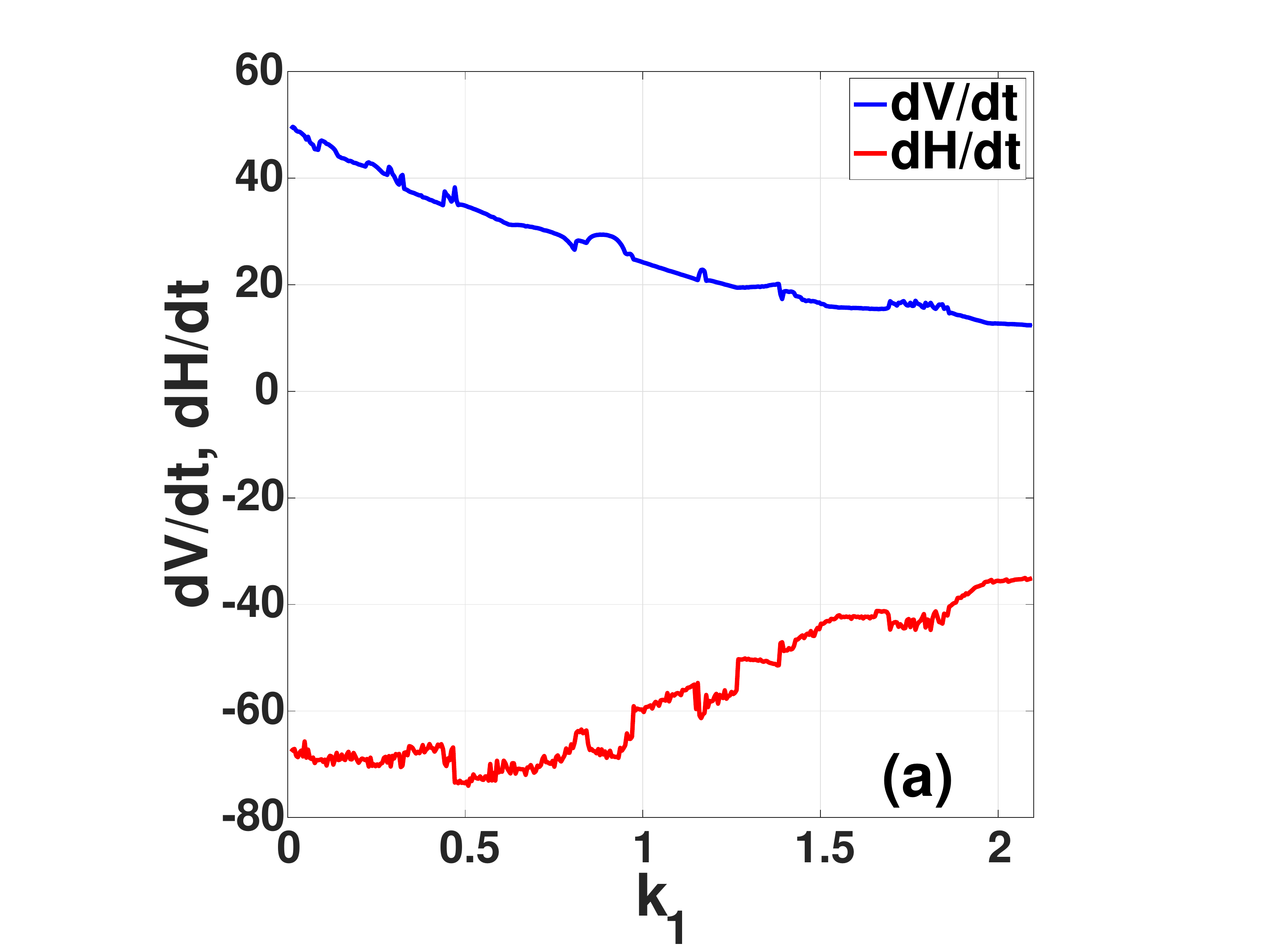}\includegraphics[width=0.25\textwidth]{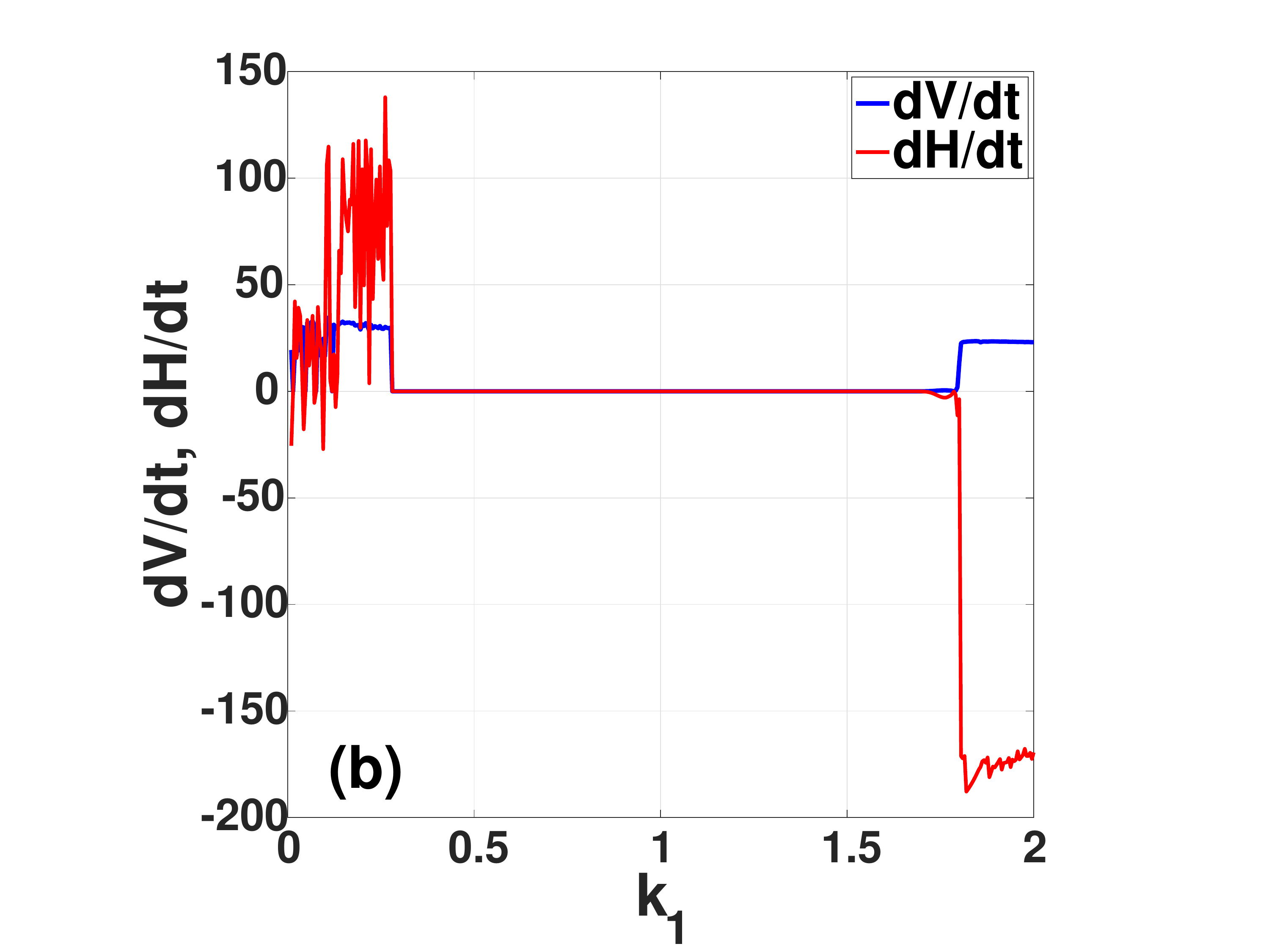}
\includegraphics[width=0.25\textwidth]{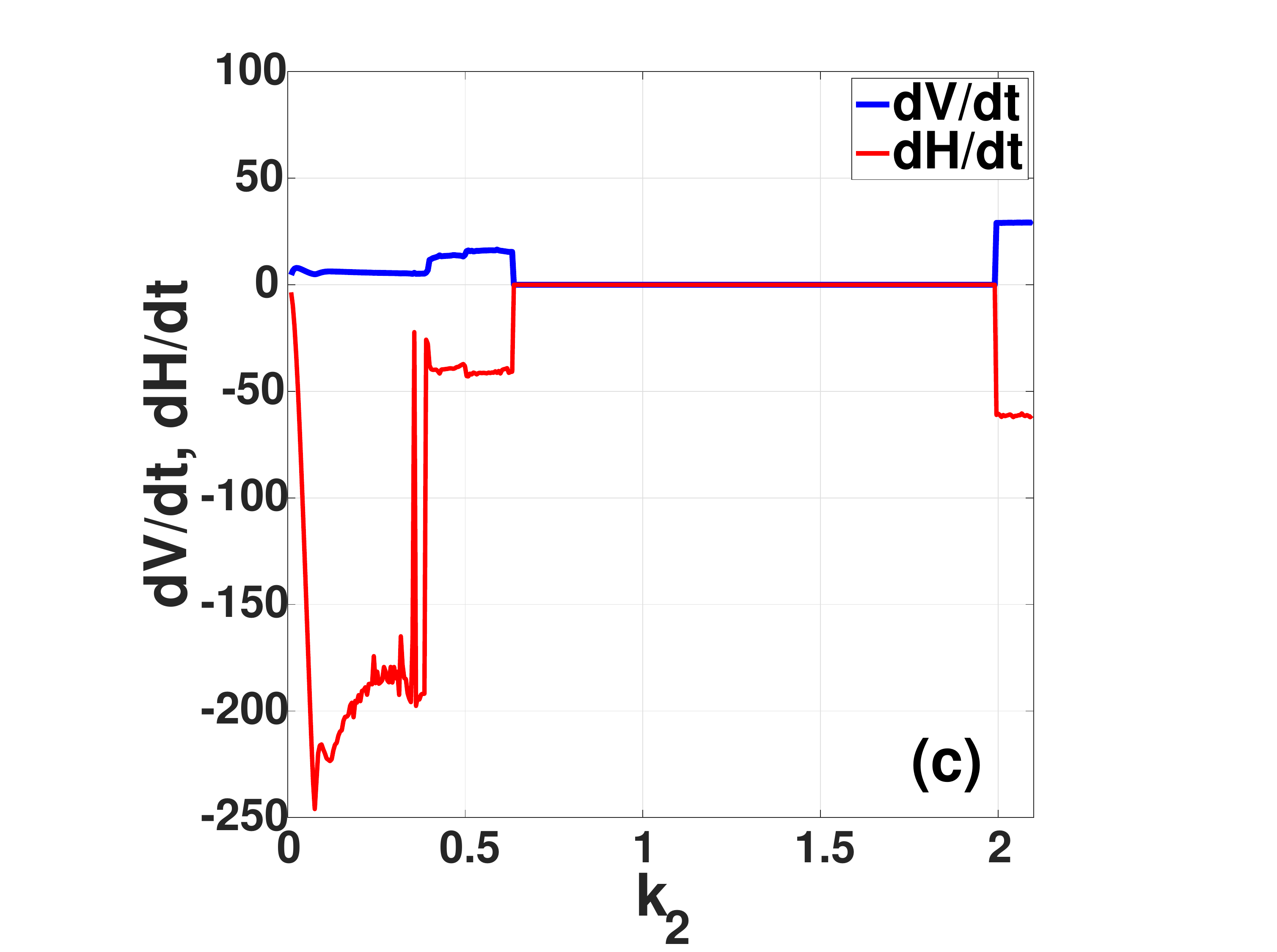}\includegraphics[width=0.25\textwidth]{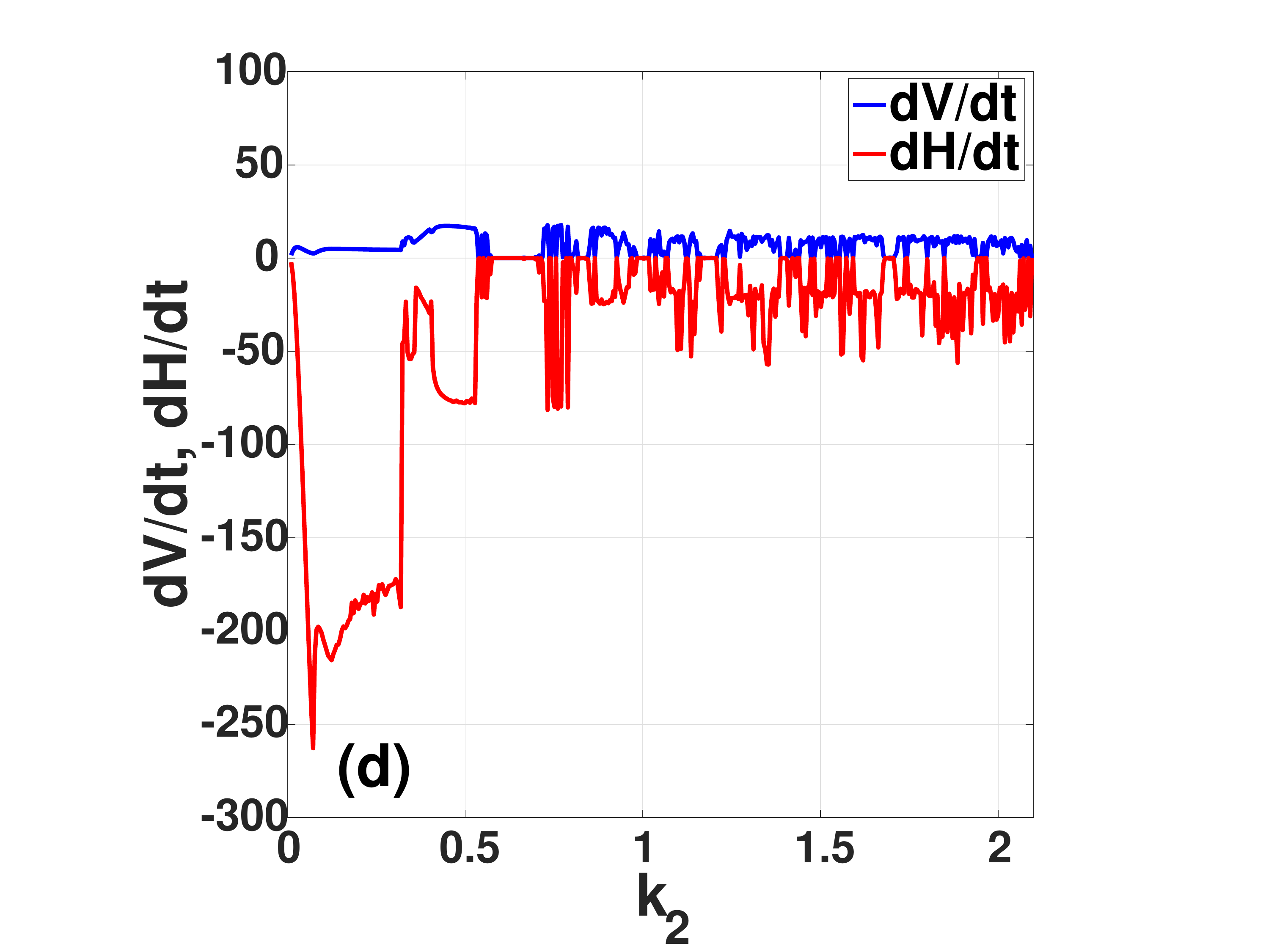}
\caption{Panels \textbf{(a)} and \textbf{(b)} show the variations $dV/dt$ and $dH/dt$ with resp. to  $k_1$.
In \textbf{(a)}: $g_e = 0.1$,  $g_c = 0.1$, $k_2 = 0.5$. Here, for all values of $k_1$,
$dV/dt>0$ indicating the instability of the synchronized state with $0<dH/dt<0$.
In \textbf{(b)}: $g_e = 2.75$, $g_c = 1.35$ $k_2 = 0.5$. Here, for intermediate values $k_1$, 
we have $dV/dt=0$ indicating a stable synchronized state with $dH/dt=0$; and for small and large values of $k_1$, $dV/dt>0$ 
indicating the instability of the synchronized state with $0<dH/dt<0$.
Panels \textbf{(c)} and \textbf{(d)} show $dV/dt$ and $dH/dt$ with resp. to  $k_2$.
In \textbf{(c)}: $g_e = 0.1$, $g_c = 0.1$,  $k_1 = 1.9$.  
In \textbf{(d)}: $g_e = 0.95$, $g_c = 0.5$  $k_1 = 1.9$.
In both panels, $dH/dt=0$ only when $dV/dt=0$, indicating a stable synchronized state, 
while $dH/dt\neq0$ only when $dV/dt>0$, indicating unstable synchronized state.}\label{fig:7}
\end{center}
\end{figure}

In Fig.\ref{fig:7}, we show the variations of $dV/dt$ and $dH/dt$ with respect to the 
memristive gain parameters $k_1$ and $k_2$ in a weak and strong synaptic coupling regimes. 
In Fig.\ref{fig:7}\textbf{(a)}, we consider weak synaptic coupling strength and we always have $dH/dt<0$ 
whenever $dV/dt>0$ for all the values of $k_1$, indicating unstable synchronized states. 
In Fig.\ref{fig:7}\textbf{(b)}, we switch to a stronger synaptic coupling strengths and we get 
stable synchronized states as $dV/dt=0$ and $dH/dt=0$ for 
intermediate values of $k_1$, and unstable synchronized states ($dV/dt>0$ and $dH/dt\neq0$) for 
small and large values of $k_1$.

In Fig.\ref{fig:7}\textbf{(c)} and \textbf{(d)}, 
we also consider $k_2$ in weak and strong synaptic coupling regimes, respectively. And again we always the same behavior, i.e., 
synchronized states for $dV/dt=0$ and $dH/dt=0$, and unstable synchronized state when $dV/dt>0$ and $dH/dt\neq0$. Especially, 
in Fig.\ref{fig:7}\textbf{(d)}, where $dV/dt$ and $dH/dt$ vary a lot, we can see that the signs of  $dV/dt$ and $dH/dt$ are 
always mirror images of each other about zero-symmetry line, i.e., $dH/dt=0$ only when $dV/dt=0$ (stable synchronized states), 
and $dH/dt<0$ only when $dV/dt>0$ (unstable synchronized states).   

\begin{figure}[htp!]
 \begin{center}
 \includegraphics[width=0.25\textwidth]{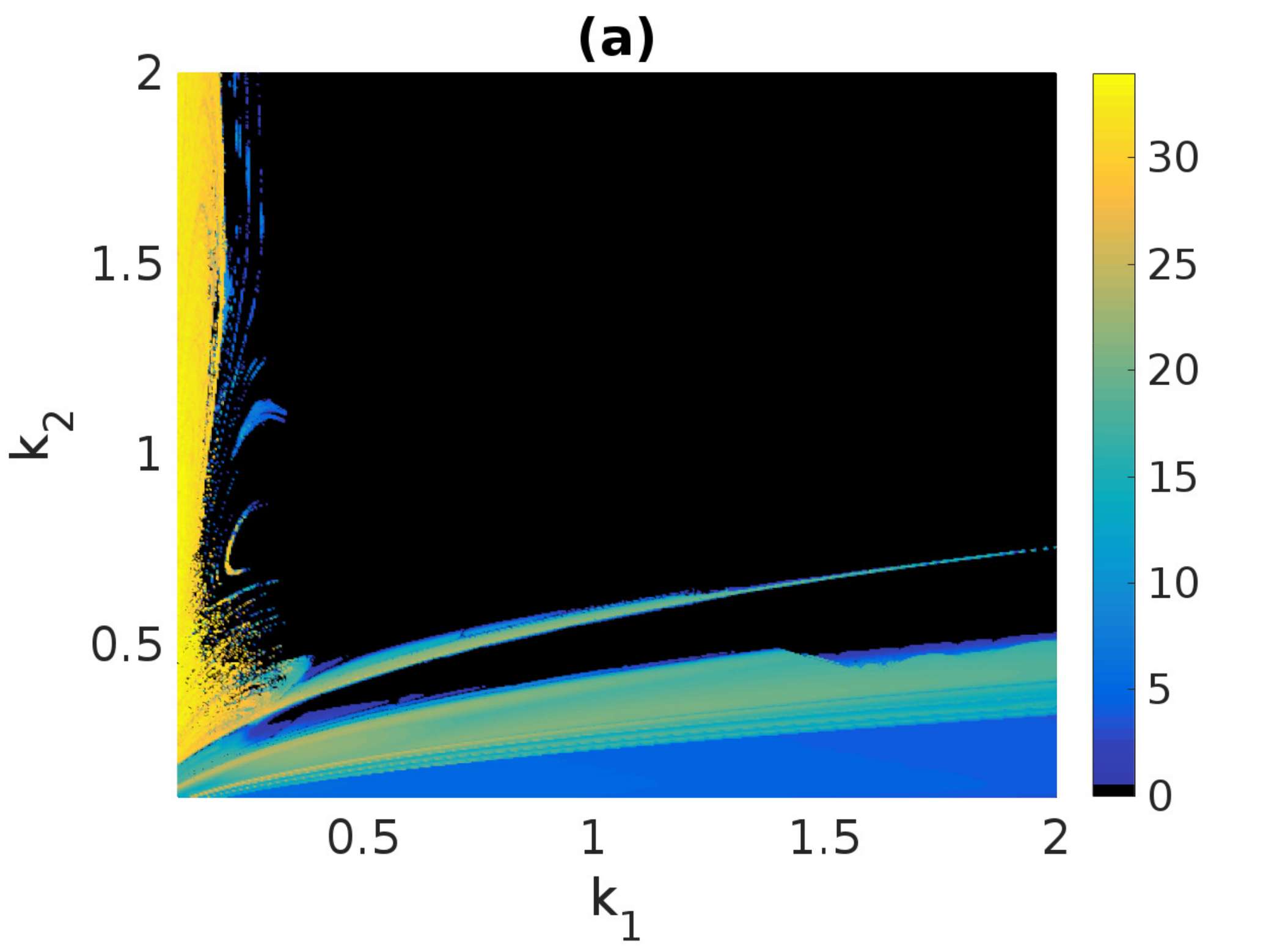}\includegraphics[width=0.25\textwidth]{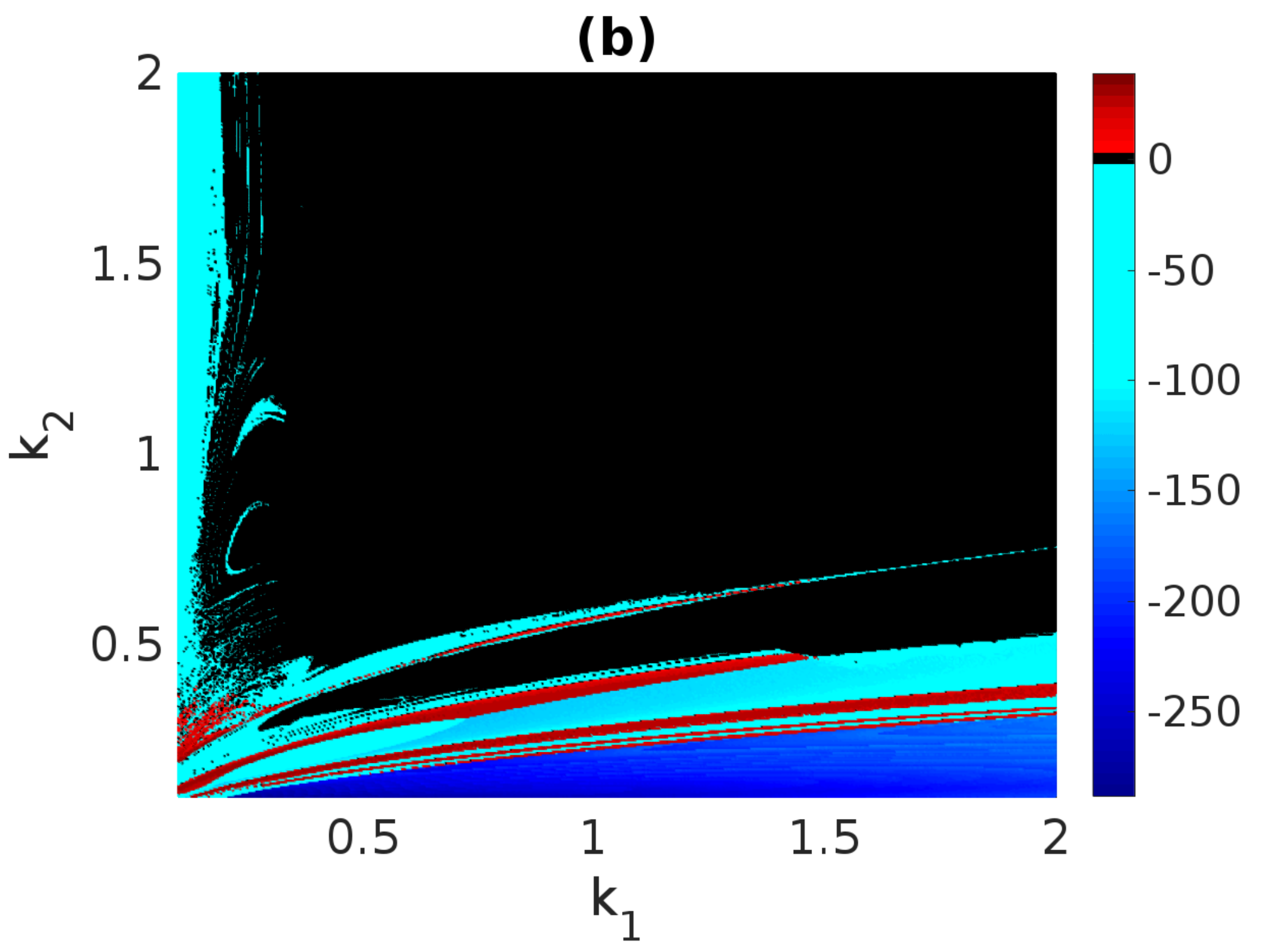}
 \includegraphics[width=0.25\textwidth]{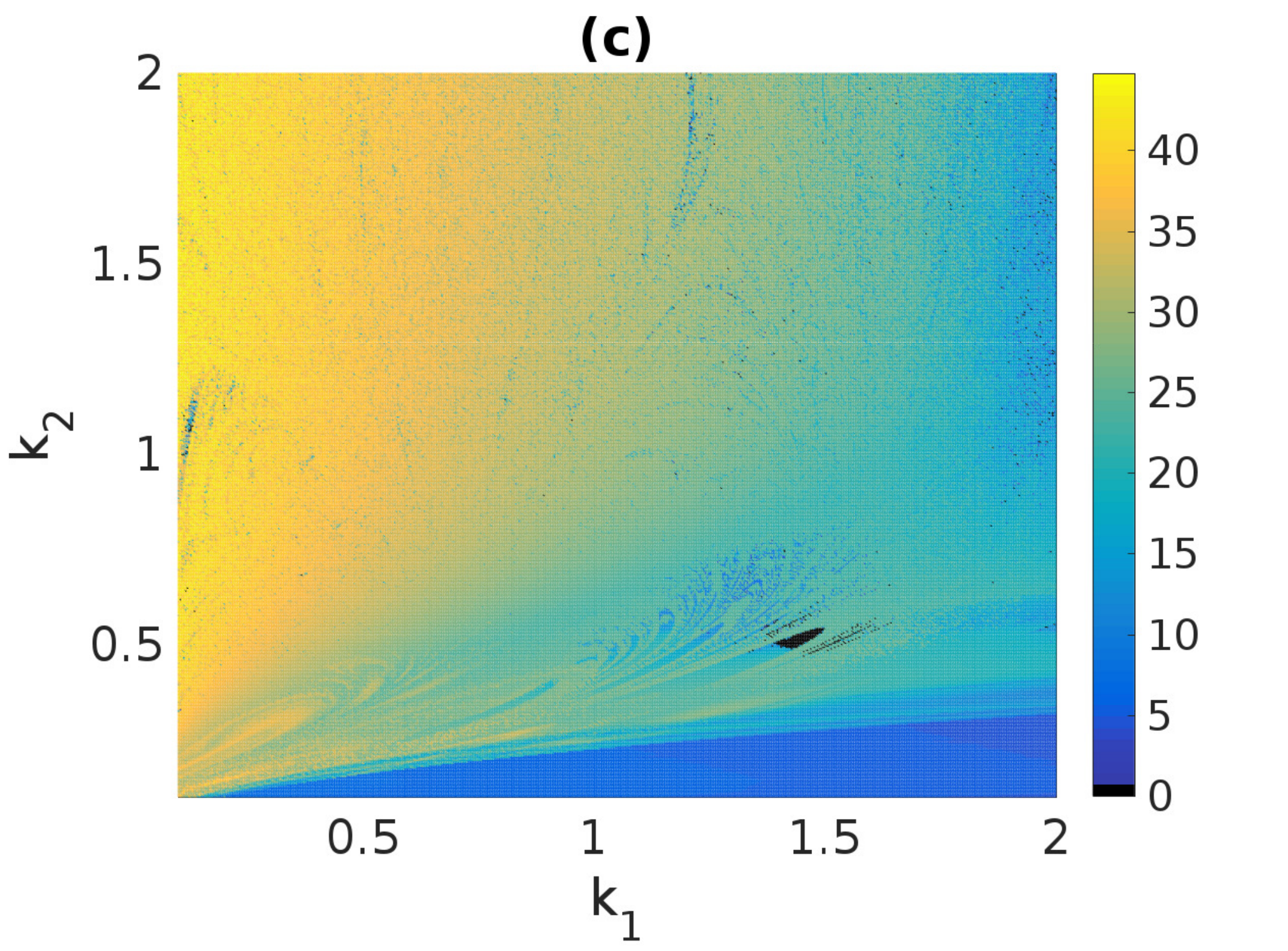}\includegraphics[width=0.25\textwidth]{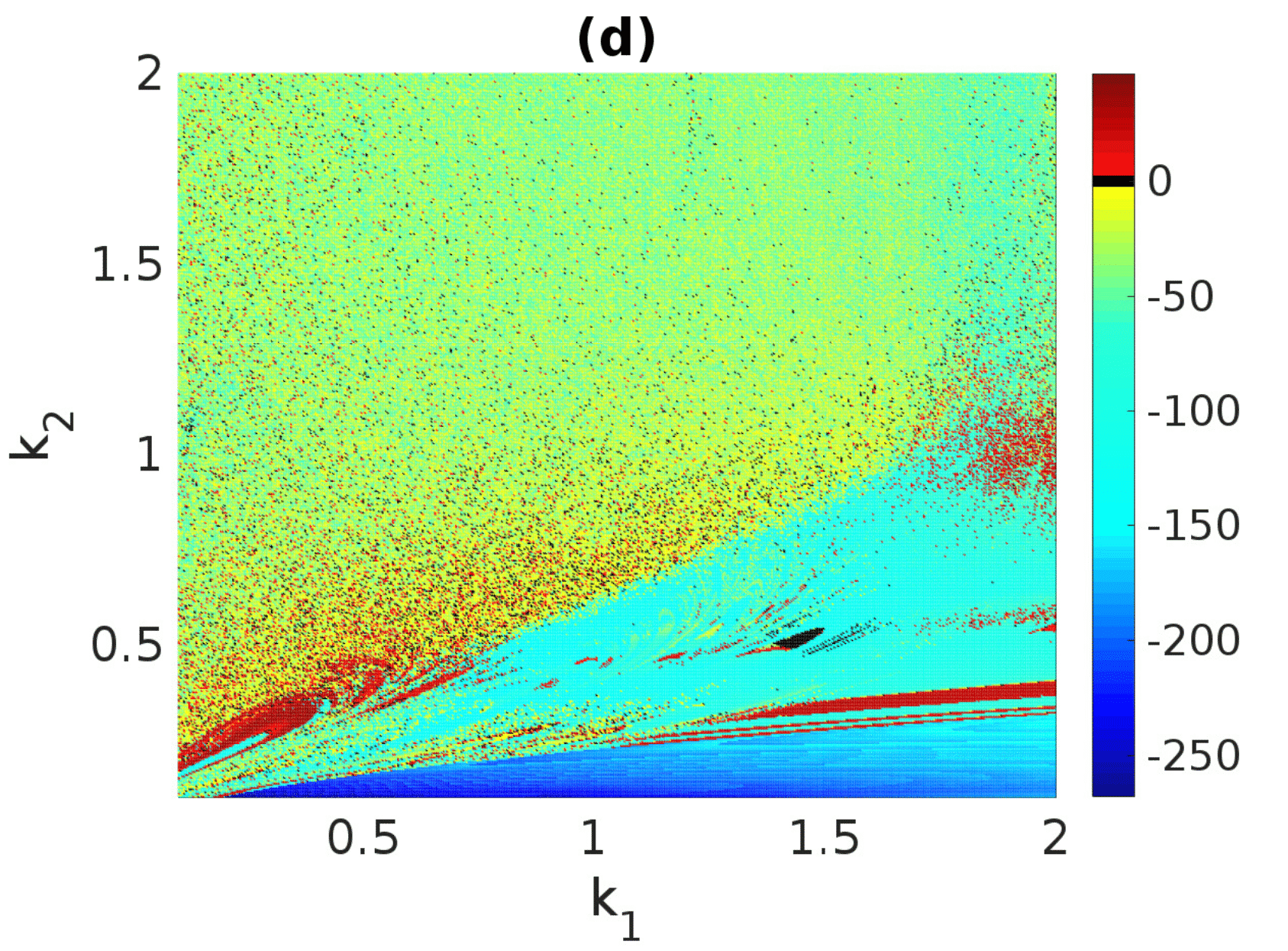}
\caption{Color-coded variations of $dV/dt$ in panel \textbf{(a)} and the corresponding $dH/dt$ in panels \textbf{(b)}, with resp. to $k_1$ and 
$k_2$ in a strong coupling regime with $g_e=4.0, g_C=2.75$. 
We see that stable synchronized states emerge at values of $k_1$ and $k_2$ for which $dV/dt>0$ and $0<dH/dt<0$, while 
the unstable synchronized state at $dV/dt=0$ and $dH/dt=0$.
In a weak coupling regime with $g_e=1.5, g_c=0.75$ panels \textbf{(c)} and \textbf{(d)} respectively show the color-coded variations of $dV/dt$ and $dH/dt$.
Here, we also have $0<dH/dt<0$ only when $dV/dt>0$ and $dH/dt=0$ (or $dH/dt\approx0$) only when $dV/dt=0$.}\label{fig:8}
 \end{center}
 \end{figure}

To have a global view on the behavior of $dV/dt$ and  $dH/dt$  with respect to the memristive parameters $k_1$ and $k_2$,
we computed these functions in a two-parameter space. Fig.\ref{fig:8}\textbf{(a)} and \textbf{(b)} show a color-coded $dV/dt$ and $dH/dt$ as a function of $k_1$ and $k_2$, respectively,
in a strong synaptic coupling regime. In Fig.\ref{fig:8}\textbf{(a)}, we observed synchronized states (the black regions) where $dV/dt=0$ corresponds to the 
black regions in Fig.\ref{fig:8}\textbf{(b)} where $dH/dt=0$ or or $dH/dt\approx0$. And wherever $dV/dt>0$ in Fig.\ref{fig:8}\textbf{(a)}, 
we have $dH/dt\neq0$ (i.e., either positive or negative), indicating unsynchronized states. 

In Fig.\ref{fig:8}\textbf{(c)} and \textbf{(d)}, we switch to a weak synaptic coupling regime. Here, panels \textbf{(c)} and \textbf{(d)} also 
show a color-coded $dV/dt$ and $dH/dt$ as a function of $k_1$ and $k_2$, respectively. Here, we mostly have $dV/dt>0$ (unsynchronized state) in regions where $dH/dt\neq0$. 
However, many black spots (where $dV/dt=dH/dt=0$, indicating synchronized states) can also be observed. For example, in panels \textbf{(c)} and \textbf{(d)}, 
around the coordinates $(k_1,k_2)=(1.5,0.5)$, a black patch of synchronized states ($dV/dt=dH/dt=0$) can be observed.  
 
A theoretical explanation for this sign correlation between time rate of change of the Lyaponuv $dV/dt$ and Hamilton $dH/dt$ functions and hence,
the ability of $dH/dt$ to also indicate whether or not a synchronized state is asymptotically stable is the following: 
First, we notice that $dH/dt$ can be either positive or negative (only when $dV/dt>0$, indicating unstable synchronized states). 
When $dV/dt>0$ , the chaotic system (i.e., the coupled neurons) initially located
outside its synchronization manifold  would gain ($dH/dt>0$) or lose energy ($dH/dt<0$) 
in its movement towards synchronization manifold, where $dV/dt=0$ (indicating an asymptotically stable synchronous state) and $dH/dt=0$ or $dH/dt\approx0$ (i.e., or vanishingly small). 
This is so because on the synchronization manifold , the trajectory will repeatedly return to arbitrarily
close states in the bounded phase space of the chaotic attractor and as a result to arbitrarily close energy values. 
Hence, on synchronization manifold the average time rate of of Hamilton energy function will be zero, i.e., $dH/dt=0$ or vanishingly small, i.e., $dH/dt\approx0$. 
This implies that, in general, all the different regimes of synchronization that the
two neurons attain at different values of the system's parameter would occur at zero (or very low) net dissipation of energy (i.e., at $dH/dt=0$ or $dH/dt\approx0$). 
But, there could be ranges of parameter values where the activity of the coupled system is more
demanding energetically, that is when $0<dH/dt<0$ which as we result from the numerical simulation occur only when $dV/dt>0$, indicating unstable synchronized states. 
This means that when the neurons are out of the synchronization manifold, a non-zero energy dissipation ($0<dH/dt<0$) is thus necessary 
to drive the coupled neurons to the synchronization manifold where the net energy dissipated become zero or very low ($dH/dt=0$ or $dH/dt\approx0$) 
and where the time rate of change of the Lyaponuv function is also zero ($dV/dt=0$) -- indicating an asymptotically stable synchronized state following Krasovskii-Lyapunov stability theory. 
Hence, the time rate of change of the Hamilton function of an error dynamical system associated to a coupled system can be used as a synchronization stability function.
%
 
%%%%%%%%%%%%%%%%%%%%%%%%%%%%%%%%%%%%%%%%%%%%%%%%%%%%%%%%%%%%%%%%%%%%%%%%%%%%%%%%%%%%%%%%%%%%%%%%%%%%%%%%%%
\section{Summary and concluding remarks}\label{sect5}
%%%%%%%%%%%%%%%%%%%%%%%%%%%%%%%%%%%%%%%%%%%%%%%%%%%%%%%%%%%%%%%%%%%%%%%%%%%%%%%%%%%%%%%%%%%%%%%%%%%%%%%%%%%
In this work, we have investigated the dynamics of an improved version of the standard 3D Hindmarsh-Rose neuron model by taking into account not only 
the exchange of calcium ions across the cell membrane, but by also considering the dynamics of the magnetic flux induced across the membrane as a result of the motion of ions.
The electric activity of the improved 5D Hindmarsh-Rose model shows upon variations of the external input current and the magnetic flux (memristive gain) parameters,
a rich dynamical behavior including periodic/chaotic spiking and bursting, chaotic super-bursting -- dynamical behaviors observed in real biological neurons upon variations of the corresponding parameters.

To investigate the synchronization dynamics of two coupled 5D  Hindmarsh-Rose neurons, we considered a pair neurons coupled via both a instantaneous electrical 
and inhibitory chemical synapses.  Using the Krasovskii-Lyapunov stability theory, we prove that the synchronization manifold of the coupled neurons can be  asymptotically
stable for suitable values of the electrical and chemical coupling strengths $g_e$ and $g_c$, and the memristive gain parameters $k_1$ and $k_2$.
Moreover, we used Helmholtz’s theorem to calculate the Hamilton energy function associated to the error dynamical of system of the coupled neurons.
Numerical computations indicated that we always have a \textit{non-zero} ($0<dH/dt<0$) time variation of the Hamilton function only when the time variation of the Lyapunov function is \textit{positive}
($dV/dt>0$), and \textit{zero} ($dH/dt=0$) (or vanishingly small, $dH/dt\approx0$) only when the time variation of the Lyapunov function is also \textit{zero} ($dV/dt=0$). 
Thus, the time variation of the Hamilton energy function of the error dynamical system associated to a coupled system,
can be used as an asymptotic stability function for synchronization. This result which might be also useful for general engineering purposes, 
paves an alternative way of determining the asymptotic stability of the 
synchronized states in coupled systems from an energy perspective without necessarily having to construct a Lyapunov function which might be a difficult task for systems modeled 
with more complicated mathematical equations.

The work presented in this paper could be extended in two directions. First, by considering synaptic and/or channel noise (e.g., Gaussian white noise or non-Gaussian colored noises), due
to their presence and relevance in real neural dynamics. And secondly, by investigating the variation of the synchronization energy between the layers 
of a multiplex neural network -- a relevant network structure ubiquitous in the brain.

\begin{acknowledgements}The author gratefully acknowledges the financial support from the 
Lehrstuhl f\"{u}r Angewandte Analysis (Alexander von Humboldt-Professur), 
Department Mathematik, Friedrich-Alexander-Universit\"{a}t Erlangen-N\"{u}rnberg, Germany.
\end{acknowledgements}
 

 \section*{Compliance with ethical standards}
 \section*{Conflict of interest} The author(s) declare that there is no conflict
of interest in relation to this article.

\setcounter{equation}{0}
\renewcommand{\theequation}{A.\arabic{equation}}

\begin{appendices}\label{Appendix}
\addcontentsline{toc}{section}{Appendix}
\section*{Appendix}

The five ordered-main sub-determinants of  matrix \textbf{M} in Eq.~\eqref{eq:1aa} are given by
\begin{equation}\label{eq:A1}
\begin{split}
\left\{\begin{array}{lcl}
\mathbf{D_1}&=& (3aJ-2b)J+k_1\alpha+g_c(1+e^{-\lambda(J-\theta_s)})^{-2},\\[0.25cm]
\mathbf{D_2}&=& \mathbf{D_1} + 4dJ  -1 - 4 d^2 J^2 ,\\[0.25cm]
\mathbf{D_3}&=& r(2ps+\mathbf{D_2})-p^2-r^2s^2,\\[0.25cm]
\mathbf{D_4}&=& \delta \mu \mathbf{D_3} -  ( \sigma-\gamma \mu )^2\big[r\mathbf{D_1} - (p - r s)^2\big],\\[0.25cm]
\mathbf{D_5}&=& k_2\mathbf{D_4} -r \big[\delta\mu - (\sigma-\gamma \mu)^2\big].
\end{array}\right.
\end{split}
\end{equation}
It is easy to see that, for the given values of $a$ ,$b$, $d$, $r$ ,$s$ ,$p$, $\mu$, $\alpha$, $\delta$, $\gamma$, $\sigma$, 
and for suitable values of $g_c$, $k_1$, and $k_2$, we can always have
\begin{eqnarray}\label{eq:A2}
\mathbf{D_i}>0, \:\: \mathbf{i}=\{1,2,3,4,5\},
\end{eqnarray}
which makes the matrix \textbf{M} positive definite.
\end{appendices}
 
\bibliographystyle{spphys}       % APS-like style for physics
\bibliography{mybibfile}   % name your BibTeX data base

\end{document}